\documentclass[aps,preprint]{revtex4}
\input{epsf.tex}
\usepackage{graphicx}
\usepackage{epsf}
\usepackage{wrapfig}
\usepackage{epsfig}

\newcommand{\be}{\begin{eqnarray}}
\newcommand{\ee}{\end{eqnarray}}
\def\brho{{\mbox{\boldmath$ \rho$}}}

\begin{document}

\title{REALISTIC CALCULATIONS OF ONE- AND TWO-HADRON EMISSION 
      PROCESSES OFF FEW- AND MANY-BODY NUCLEI\footnote{Presented at the $6^{th}$ International Workshop on "Electromagnetically 
      Induced 
      Two-Hadron Emission", ~Pavia, September 24-27, 2003}}
\author{M. Alvioli}
\author{C.Ciofi degli Atti}
\author{L.P. Kaptari\footnote{On leave from  Bogoliubov Lab. 
      Theor. Phys.,141980, JINR,  Dubna, Russia}}
\affiliation{Department of Physics, University of Perugia\\ and 
      Istituto Nazionale di Fisica Nucleare, Sezione di Perugia, 
      Via A. Pascoli, I-06100, Italy}
\author{H. Morita}
\affiliation{Sapporo Gakuin University, Bunkyo-dai 11, Ebetsu 069,
      Hokkaido, Japan}

\begin{abstract}
The exclusive electro-disintegration processes $^2H(e,e^\prime p)n$,
$^3He(e,e^\prime p)^2H$, $^3He(e,e^\prime p)pn$,  $^3He(e,e^\prime p p)n$
and $^4He(e,e^\prime p)^3H$
have been calculated using realistic few-body wave functions and treating
final state interaction (FSI) effects within a generalized eikonal approach.
The semi-inclusive scattering $A(e,e^\prime p)X$ off complex nuclei has 
been analyzed using many-body wave functions for $^{16}O$ and $^{40}Ca$,
obtained within the framework of a linked cluster expansion and taking FSI 
into account by a Glauber-type approach.
The effect of color transparency has  also  been included by
considering the Finite Formation Time (FFT) that the hit hadron needs to reach
its asymptotic physical state.
\end{abstract}
\maketitle
\setcounter{page}{1}

%
%

\section{Introduction}

\noindent
One of the main aims of  nowadays hadronic physics is the investigation of the
limits  of validity  of the so called \textit{Standard Model} of nuclei, i.e.
the description of nuclei in terms of the solution of the non relativistic
Schr\"odinger equation containing realistic nucleon-nucleon interactions.
To this end, exclusive lepton scattering could be very useful for it might yield relevant information
on the nuclear wave function,
provided  the initial and final 
states involved in the scattering process are described within a consistent, reliable approach.
In the case of 
\textit{few-body systems},  a consistent treatment of initial and final
states is nowadays possible at low energies (see e.g. \cite{gloeckle,pisa} 
and References therein quoted), but at high energies,  when the number of 
partial waves sharply increases  and  nucleon excitations can occur, the 
Schr\"odinger approach  becomes impractical and other methods have to be employed.
In the case of \textit{complex nuclei}, additional difficulties arise 
due to the approximations which are still necessary to solve the many-body problem.
As a matter of fact, in spite of the fundamental progress made in recent years in the 
calculation of the properties of light nuclei (see e.g. \cite{wiri}), much remains  to be done, 
also in view that the results of very sophisticated calculations
(\textit{e.g.} the variational Monte Carlo ones \cite{pie01}), show that 
the wave function which minimizes the expectation value of the Hamiltonian,
provides a very poor nuclear density; moreover, the structure of the best trial 
 wave function is so complicated, that its application to the 
calculation of various processes at intermediate and high energies, where the 
role of the  so called \textit{nuclear effects} is becoming more and more 
visible, appears to be not easy task. The aim of this talk is to summarize the activity
carried out by the Perugia group in the field of the theoretical treatment of exclusive
and semi-exclusive lepton scattering off both few- and many-nucleon systems. In the former case,
{\it exact}  realistic few-body wave functions have been  used, whereas in the latter case, {\it reasonable} 
realistic many-body wave functions obtained from a cluster expansion calculation of the ground state energy,
have been employed; these wave functions, which  explain  {\it reasonably} well the ground state energy,
 density and momentum
distributions of complex nuclei, have, at the same time, a structure such that their application
to various scattering problems is rather straightforward. The structure of the paper is as follows:
in Section \ref{sec:2} few-body nuclei ($^2H$, $^3He$ and $^4He$) are discussed
giving in Section \ref{subs:2a} the basic formulae
for the calculations of the
exclusive  $A(e,e^\prime p)B$ and $A(e,e^\prime p p)C$ processes; 
in Section \ref{subs:2b} the numerical results for the processes 
$^2H(e,e^\prime p)n$,
$^3He(e,e^\prime p)^2H$, $^3He(e,e^\prime p)(p n)$ and $^4He(e,e^\prime p)^3H$,
 treating the 
effects of the final state interaction (FSI) by a generalized eikonal approach,
and  also considering color transparency effects, are presented. In Section \ref{sec:3}
 complex nuclei are discussed: the basic formalism  of the cluster expansion technique  is illustrated in 
Section \ref{subs:3a} where the results of calculations for the ground-state energy, density and momentum distributions 
are presented; 
the effects of FSI in $(e,e^\prime p)$ processes off complex nuclei is discussed in 
Section  \ref{subs:3b} in 
terms of a generalized Glauber approach; the color transparency effects are 
introduced in Section  \ref{subs:3c}; eventually, 
in Section  \ref{sec:4} the Conclusion are drawn.

\section{Few-body nuclei (${\bf Refs.}$ \cite{greno,w,claleo,mor01,misha,cio01,bra01})}
\label{sec:2}

\subsection{Basic formulae}
 \label{subs:2a}
\vskip 2mm
\begin{figure}[!htp] 
\hspace*{-2mm}                     
\epsfig{file=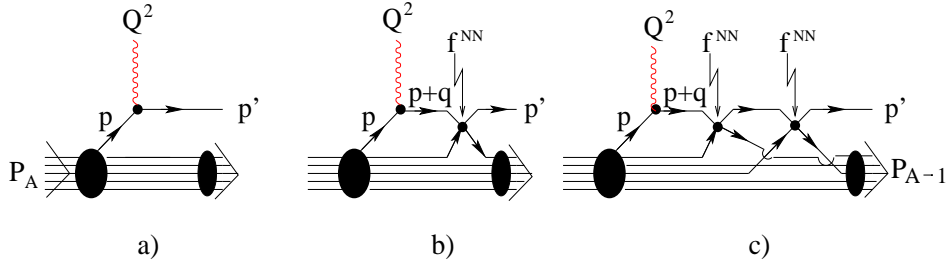,width=12.5cm,height=4.5cm}
\caption{ The Feynman diagrams for the process $A(e,e^\prime p)(A-1)$: the  Plane Wave 
      Impulse Approximation (PWIA) $a)$, and the single $b)$  and double $c)$
      rescattering in the final state. $f^{NN}$ denotes the elastic
      nucleon-nucleon (NN) scattering amplitude.}\label{fig1}
\end{figure}

In the one-photon-exchange approximation we write the differential cross
section of the process $A(e,e^\prime p)(A-1)$ in the following form 

\be
&&
\frac{d^6\sigma}{dE_{e^\prime}d\Omega_{e^\prime}d{\bf p}_m}=
      K(x,Q^2,{\bf p}_m)\, \sigma_{cc1}^{eN}(Q^2,{\bf p}_m)
      \,\left|T_{A,A-1}({\bf p}_m,E_{m})\right|^2\,,
\label{eq2}
\ee
where
$K(x,Q^2,{\bf p}_m)$ is  a kinematical factor,
$\sigma_{cc1}^{eN}(Q^2,{\bf p}_m)$  the De Forest CC1 cross section \cite{forest},
${\bf p}_m\equiv {\bf q}-{\bf p}^\prime$ the  \textit{missing momentum},  i.e.  the
Center-of-Mass momentum of the undetected particles,~ ${\bf p}^\prime$ the momentum
of the detected proton, and
$E_{m} = \sqrt{P_{A-1}^2}+M_N-M_{A} = q_0-T_{p^\prime} - T_{A-1}$  the
\textit{missing} (or \textit {removal}) \textit{energy}.

In our approach the nuclear transition matrix element
$T_{A,A-1}({\bf p}_m,E_{m})$ in eq. (\ref{eq2}) is computed 
by evaluating the corresponding Feynman diagrams of Fig. \ref{fig1},
describing the interaction of the incident electron with one nucleon of the
target  followed by its elastic  rescattering with the   nucleons of the $(A-1)$ 
nucleus. 
 
\subsection{ Results of calculations}
 \label{subs:2b}
\vskip 4mm

\centerline{\textbf{The process  $^2H(e,e^\prime p)n$}}
\vskip 4mm  

\begin{figure}[!htp]                     
\begin{minipage}{6cm}
\epsfig{file=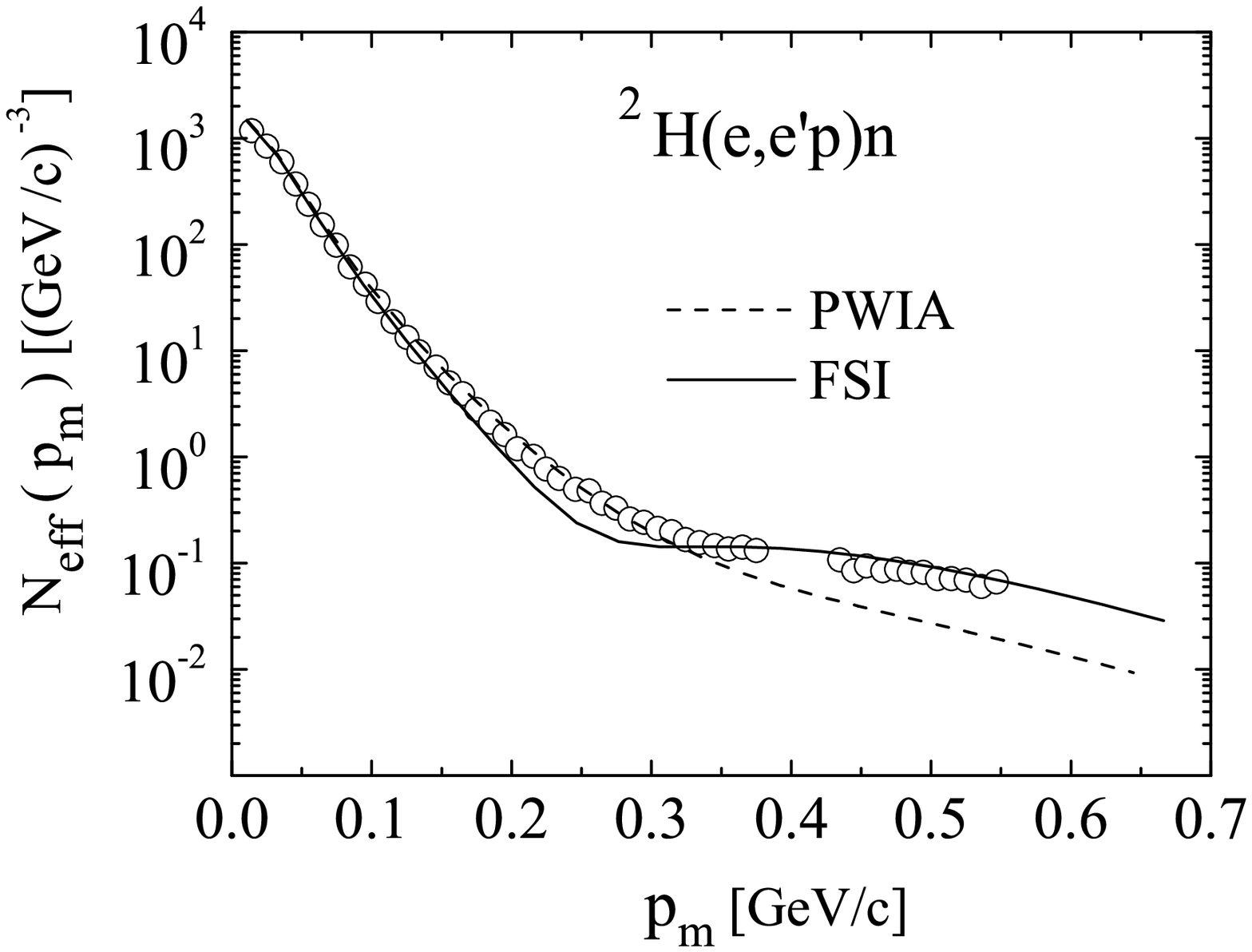,width=5.60cm,height=4.80cm}
\end{minipage}
\begin{minipage}{6cm}
\vskip 1mm
\epsfig{file=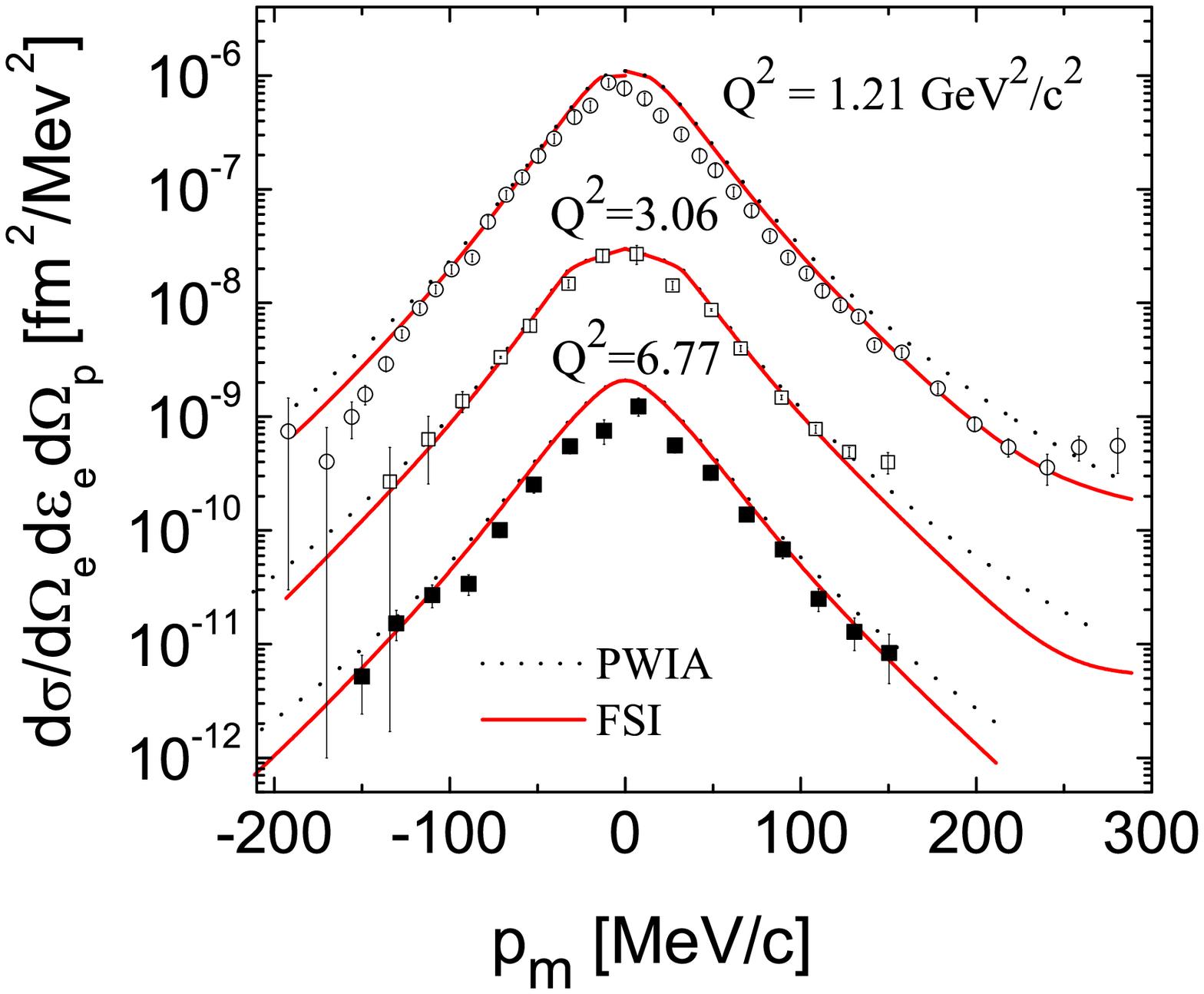,width=5.60cm,height=4.8cm}
\end{minipage}
\caption{The process $^2H(e,e'p)n$: comparison between  theoretical calculations \protect\cite{greno}
 with the experimental data 
      from JLAB \protect\cite{ulmer} (left)  and SLAC \protect\cite{bulten} 
      (right). The negative values of \protect $p_{m}$ correspond to protons
      detected at $\phi=0$. (After \cite{greno}).}
      \label{fig2}
\end{figure}

In PWIA the square of the transition matrix element in eq. (\ref{eq2}) simply 
becomes  the deuteron momentum
distribution, i.e.
  
\be
\left|T_{A,A-1}\right|^2 \to n_{D}( |{\bf p}_m|) =
\frac13\frac{1}{(2\pi)^3} \sum\limits_{{\cal
M}_D} \left | \int\, d  {\bf r} \Psi_{{1,\cal
M}_D}( {\bf r})  \chi_f\,\exp (-i{\bf p}_m {\bf r}) \right |^2
\label{deut}
\ee
where $\Psi_{{1,\cal
M}_D}( {\bf r})$ is the realistic (containing $S$ and $D$ waves) deuteron wave function, 
and the missing momentum  ${\bf p}_m =-{\bf p}$  is nothing but the inverse of the 
momentum of the bound proton (cf. Fig 1a)).  When the  FSI  is taken into account, ${\bf p}_m \neq -{\bf p}$ 
and the final state of the ($n p$) pair should be  described by the  solution of
the Schr\"odinger equation  in the continuum.
When the relative energy of the  ($n p$) pair is 
large, the two-nucleon continuum wave function can be approximated by its eikonal  form, obtaining 
\cite{w}
\begin{equation}
n_{D} \to N_{eff}({\bf p}_m) =
\frac13\frac{1}{(2\pi)^3} \sum\limits_{{\cal
M}_D} \left | \int\, d  {\bf r} \Psi_{{1,\cal
M}_D}( {\bf r}) S( {\bf r}) \chi_f\,\exp (-i
{\bf p}_m {\bf r}) \right |^2, \label{ddistr}
\end{equation}
where
$
{\cal S}({\bf r}) = \left[ 1-\theta(z)\Gamma({\bf b})\right ]
$
, with $z$ and ${\bf b}$ being  the longitudinal and transverse co-ordinates  with
respect to the direction of the struck nucleon, describes the FSI.
In Fig.~\ref{fig2}  (left panel)  the experimental $N_{eff}$ \cite{ulmer}, {\it viz}
\begin{equation}
 N_{eff}({\bf p}_m)=
[\frac{d^6\sigma}{dE_{e^\prime}d\Omega_{e^\prime}d{\bf p}_m}]^{exp}
\cdot[
      K(x,Q^2,{\bf p}_m)\, \sigma_{cc1}^{eN}(Q^2,{\bf p}_m)]^{-1}
            \end{equation}
     is compared with the results of theoretical calculations  \cite{greno}  obtained using (as in all other calculations
     described in this paper)
$\Gamma({\bf b}) =\displaystyle{\sigma_{NN}^{tot}[(1-i\alpha)}/(4\pi b_0^2)]
exp(-{\bf b}^2/2b_0^2)$.
In the  right panel of Fig. \ref{fig2}  the $Q^2$ dependence of the
cross section is illustrated for  two different values of  the azimuthal angle $\phi$ between
the scattering and reaction planes, namely $\phi=0$ (negative values of
${\bf p}_{m}$) and $\phi=\pi$ (positive values of ${\bf p}_{m}$).      
It can be  seen that FSI effects lead, in general,  to a better agreement with
the experimental data.

\centerline{\textbf{The processes  $^3He(e,e^\prime p)^2H$ and 
      $^3He(e,e^\prime p)pn$}}
\vskip 2mm

In Ref. \cite{greno} these processes have been calculated by considering the following three cases:
\\
 \noindent  
  1. {\it the PWA approximation}: all particles in the final states are described by 
  plane waves, which means that  the transition matrix element is nothing but the three-body ground state wave function 
  in momentum space;
  
 \noindent
  2. {\it the PWIA approximation}: in this picture (Fig.\ref{fig1}a)) the struck proton is always 
  described by a plane wave and the FSI is taken into account only  in the $($n p$)$ pair
  of the  three-body channel process
  $^3He(e,e^\prime p)(np)$; 
 in these calculations calculations both the two- and three-body  wave functions
 correspond to the $AV18$ interaction \cite{av18}, with the three-body wave function from \cite{pisa};\\ 
\noindent
3. {\it the full FSI}: the  
$($n p$)$ system (ground or continuum states) is still described by  the exact
solution of the Schr\"odinger equation, whereas the  interaction of the 
 struck nucleon with the pair is treated  
 by evaluating the Feynman diagrams of Figs. 1 b) and 1c) within the 
 eikonal approximation.
For the three-body channel, one obtains
 \be 
     \label{Hepnn}
      && |T_{A,A-1}|^2 \equiv P_D({\bf p}_m, E_m) = \int d {\bf k}\\\nonumber 
 &&\left | \int\, d {\bf r} d {\brho} \phantom{\frac12}     \!\!
   \Psi_{^3He} ({\bf r}, {\brho}) {\cal S}^{FSI}(\brho,{\bf r})  
   \exp (i{\bf p}_m \brho)  \ \phi_{12}^{{\bf k}}({\bf r})
     \right |^2 \delta\left( E_{m}-E_{3}-\frac{{\bf k}^2}{M_N}\right)
     \ee 
where 
\begin{equation}
{\cal S}^{FSI}({\bf r}_1,{\bf r}_2,{\bf r}_3) =
\prod\limits_{i=1}^{2}\ \left[ 1-\theta(z_i-z_3)\, 
\rm e^{i{\mbox{$\Delta_0(z_i-z_3) $}}}\Gamma({\bf b}_i-{\bf b}_3)\right ],
\label{esse}
\end{equation}

\begin{wrapfigure}[18]{l}[0pt]{8cm}           
 \includegraphics[height=8cm,width=7.0cm]{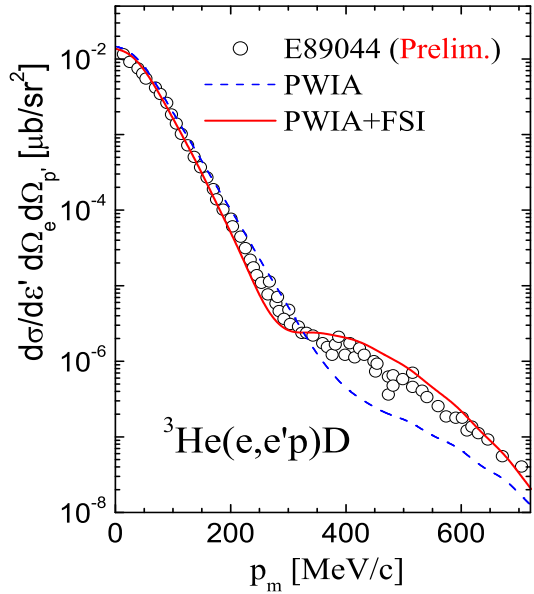}
\caption{Comparison of theoretical  calculations \protect\cite{greno} of the
      two-body channel process  with preliminary experimental data from 
      \protect\cite{jlab}. AV18 interaction \protect\cite{av18}. Three-body wave function from~\protect\cite{pisa}. 
      (After \cite{greno}).}\label{fig3}
\end{wrapfigure}
In Eq. \ref{esse} $E_3$ is the three-body threshold energy,  
and $\Delta_0\sim (q_0/|{\bf q}|) E_{m}$  a  
factor which arises when the frozen approximation underlying 
the Glauber approach is released and the recoil momentum of the third nucleon, appearing when the struck
nucleon rescatters on the second one, is taken into account\,\,\cite{glauber}, \cite{misha}. The effects from the factor
$\Delta_0$ increase with the removal energy,  but in most cases considered  they do not appreciably distort the Glauber result
(this point is still under investigation \cite{misha}).
The transition matrix element 
for the two-body channel has the same form, with the continuum two-body wave function replaced by the deuteron
wave function, and the argument of  the energy-conserving  $\delta$-function
 properly modified. In eq. \ref{Hepnn},  $P_D({\bf p}_m, E_m)$ 
 represents the  {\it distorted}
  Spectral Function, which, 
  when $\Gamma = \Delta_0 = 0$, reduces to the usual one 
 $P(k, E_{rel})$  \cite{claleo},  where  $E_{rel}=E_m-E_3$ is the relative energy of
 the ($n p$)-pair in the continuum and ${k}\equiv |{\bf p}|$  the momentum of the third nucleon.

\begin{figure}[!hpt]                   
    \includegraphics[height=0.35\textheight]{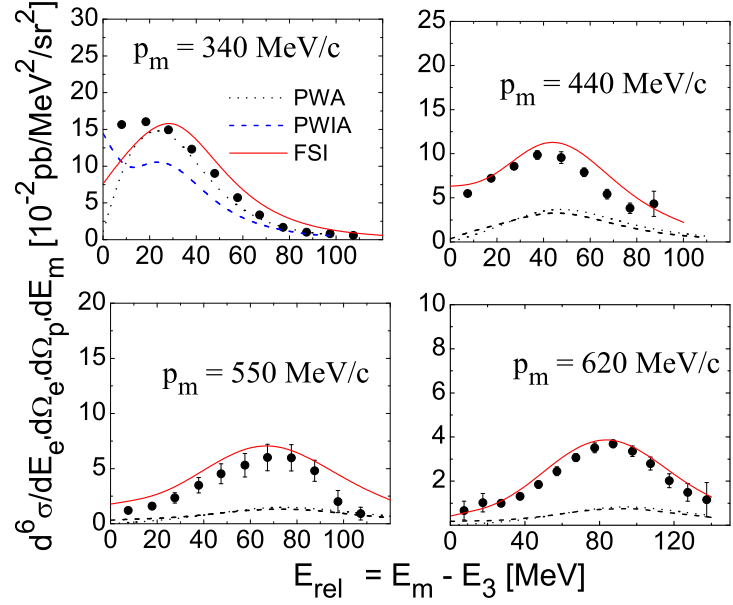}
\caption{Comparison of the theoretical calculations \protect\cite{greno}  of the process
      $^3He(e,e^\prime p)np$ with preliminary experimental results from  
      \protect\cite{jlab}. AV18 interaction \cite{av18}. Three-body wave function from \protect\cite{pisa}.
      (After \cite{greno}).}\label{fig4}
\end{figure}

The results of our calculations are shown in Figs. \ref{fig3} and  \ref{fig4}.
Both sets of data refer to the {\it  perpendicular kinematics}, when
the final  proton is detected  almost perpendicularly to  ${\bf p}_m$; it can be seen that  in the two-body channel process,  the inclusion of 
FSI effects  appreciably improves the agreement with the experimental data. 

\begin{figure}[!htp]                    
\begin{minipage}{8cm}
   \includegraphics[width=8cm,height=7.5cm]{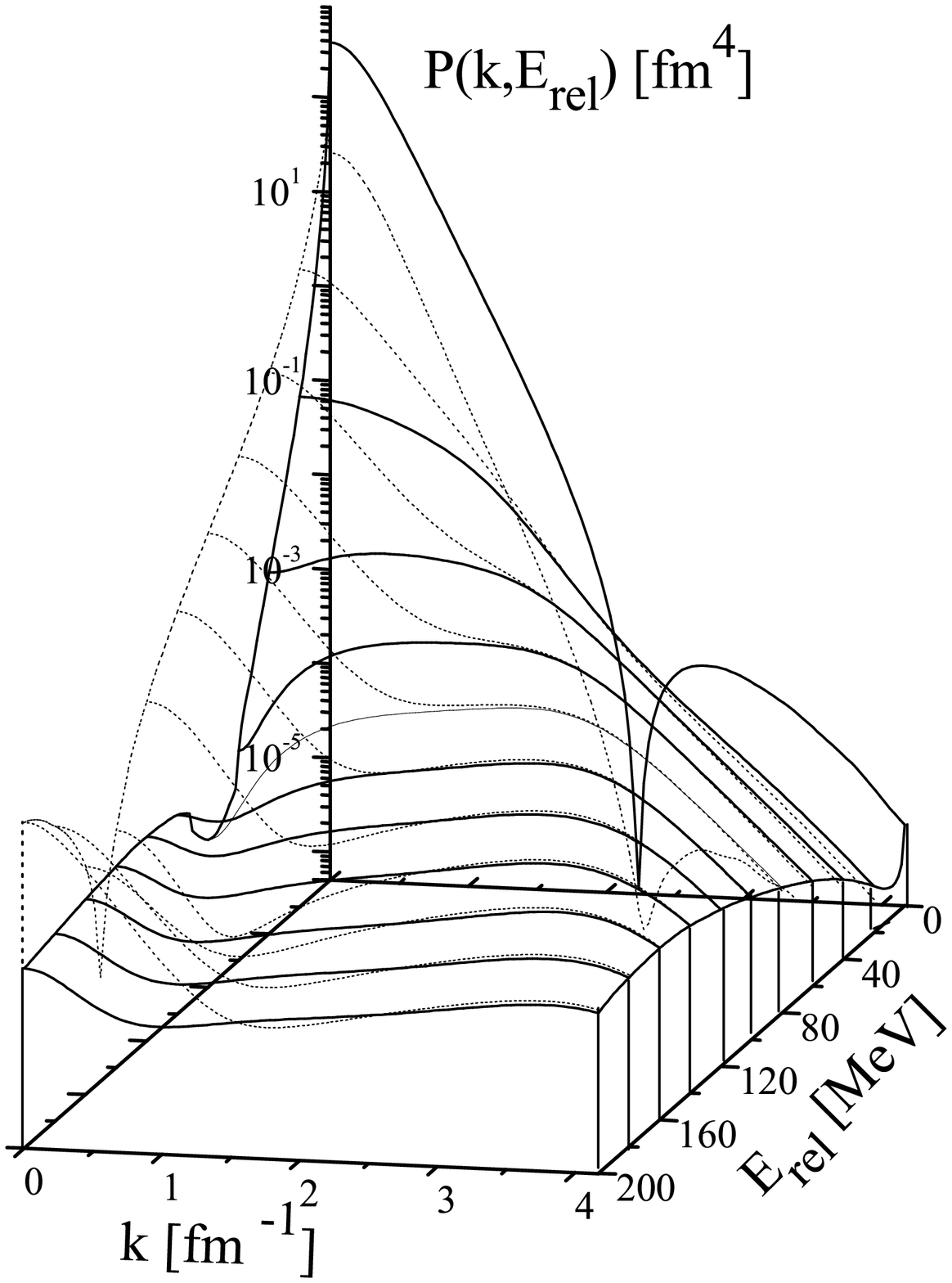}
   \end{minipage}
\begin{minipage}{7cm}
      \includegraphics[width=7cm,height=7.5cm]{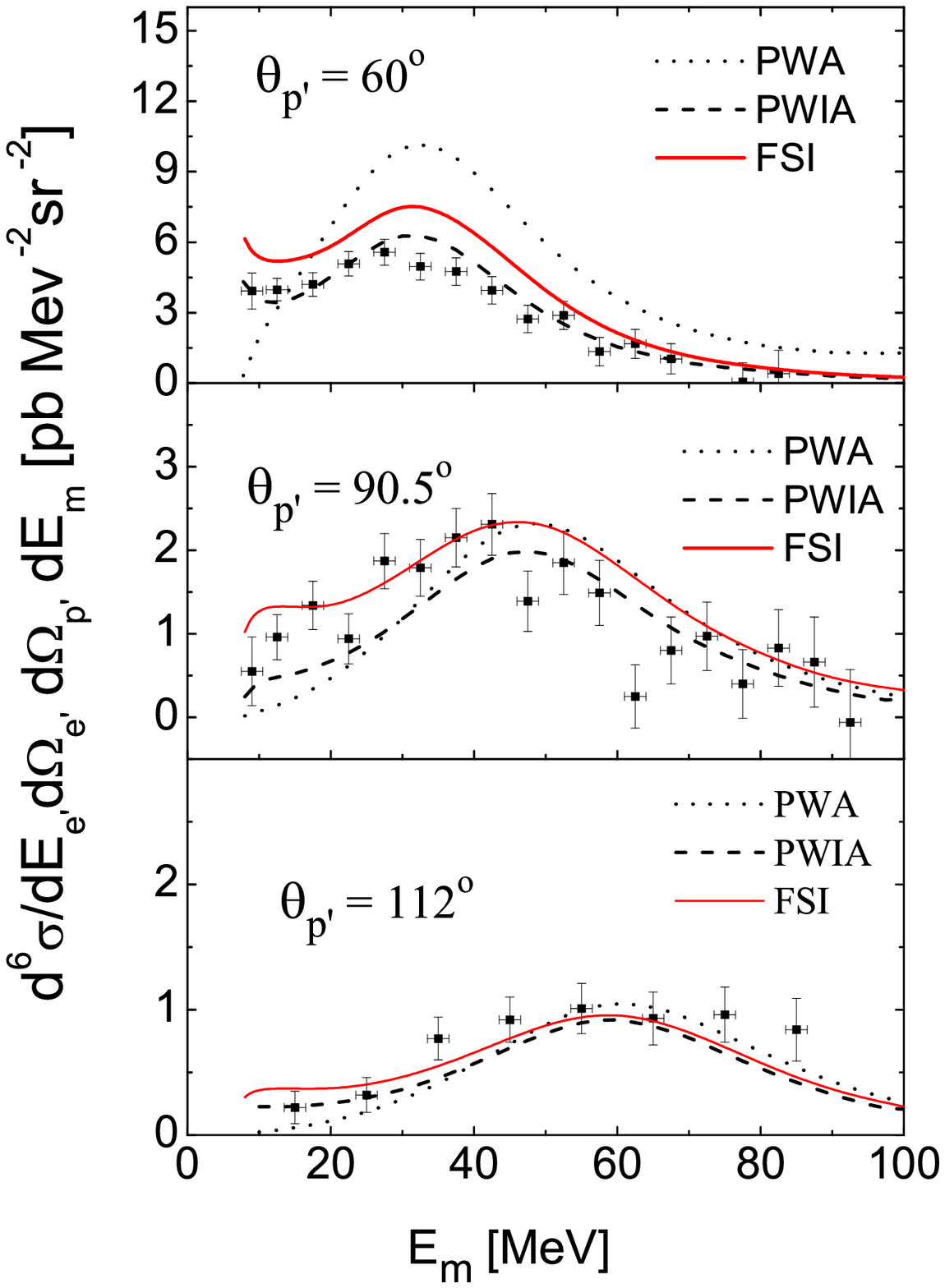}
   \end{minipage}
\caption{ {\it Left panel}: the neutron Spectral Function \cite{claleo} 
      of $^3He$ obtained with the wave functions of Ref. \cite{pisa}
      corresponding to the AV18 interaction \cite{av18}. $E_{rel}$ is
       the relative energy 
      of the (pp)-pair and  $k \equiv |{\bf p}| \equiv k_n$ the  momentum of the bound neutron.
      \textit{Right panel}: comparison of 
      our theoretical calculations (AV18 interaction \cite{av18}) of the process  
      $^3He(e,e^\prime p)np$ with the  results from  \protect\cite{saclay}. 
      Three-body wave function from \protect\cite{pisa}. (After \cite{greno}).} 
\label{fig5} 
\end{figure}
As for the three-body channel,
one sees from Fig. \ref{fig4} that  at sufficiently high values of $E_{rel}$ and $p_m$, the $PWA$ and 
$PWIA$ predictions practically
coincide, in agreement with the behaviour of the Spectral 
Function which, as shown in Fig. \ref{fig5},
exhibits bumps at $E_m \simeq {\bf k}^2/(4M_N)$ originating from two-nucleon correlations. Thus, if the PWIA
were valid, the $^3He(e,e^\prime p)(np)$ cross section at $p_m \geq 440 MeV/c$ and  $E_m \geq 10 MeV$ would be directly related
 to the
three-body wave function. Unfortunately, one sees  that in the perpendicular kinematics of~\cite{jlab},  the FSI between 
the struck proton and the $(n p)$ pair almost entirely exhausts the cross section. However, as shown in Fig. 5, this does not
seem to be the case for the experimental data of \cite{saclay}, where  the struck nucleon is detected
almost along the direction of \, ${\bf p}_m$.\\

\centerline{\textbf{The process  $^3He(e,e^\prime p p)n$}}
Extensive theoretical and experimental studies on the  $A(e,e^\prime pp)X$ process
off complex nuclei have been performed (see e.g. \cite{grab,GP,ryck}, 
and References therein quoted) and the reaction $^3He(e,e^\prime pp)n$ has been
investigated at NIKHEF \cite{groep} and  Jlab \cite{larry}. 
The effects of FSI on this process have been theoretically evaluated in \cite{glok}
and  \cite{claleo}. In the latter work, the same framework used for the $A(e,e^\prime p)B$
reaction discussed previously has been adopted,  and  the process has been considered in which $\gamma^\star$
is absorbed by the neutron in $^3He$ and the two protons are emitted by
momentum conservation. All of the three particles in the final state are allowed
to interact and  the three-body final state is described by the following
function (spin and isospin variables are omitted for ease of presentation)
\begin{figure}
  \includegraphics[height=8cm,width=7cm]{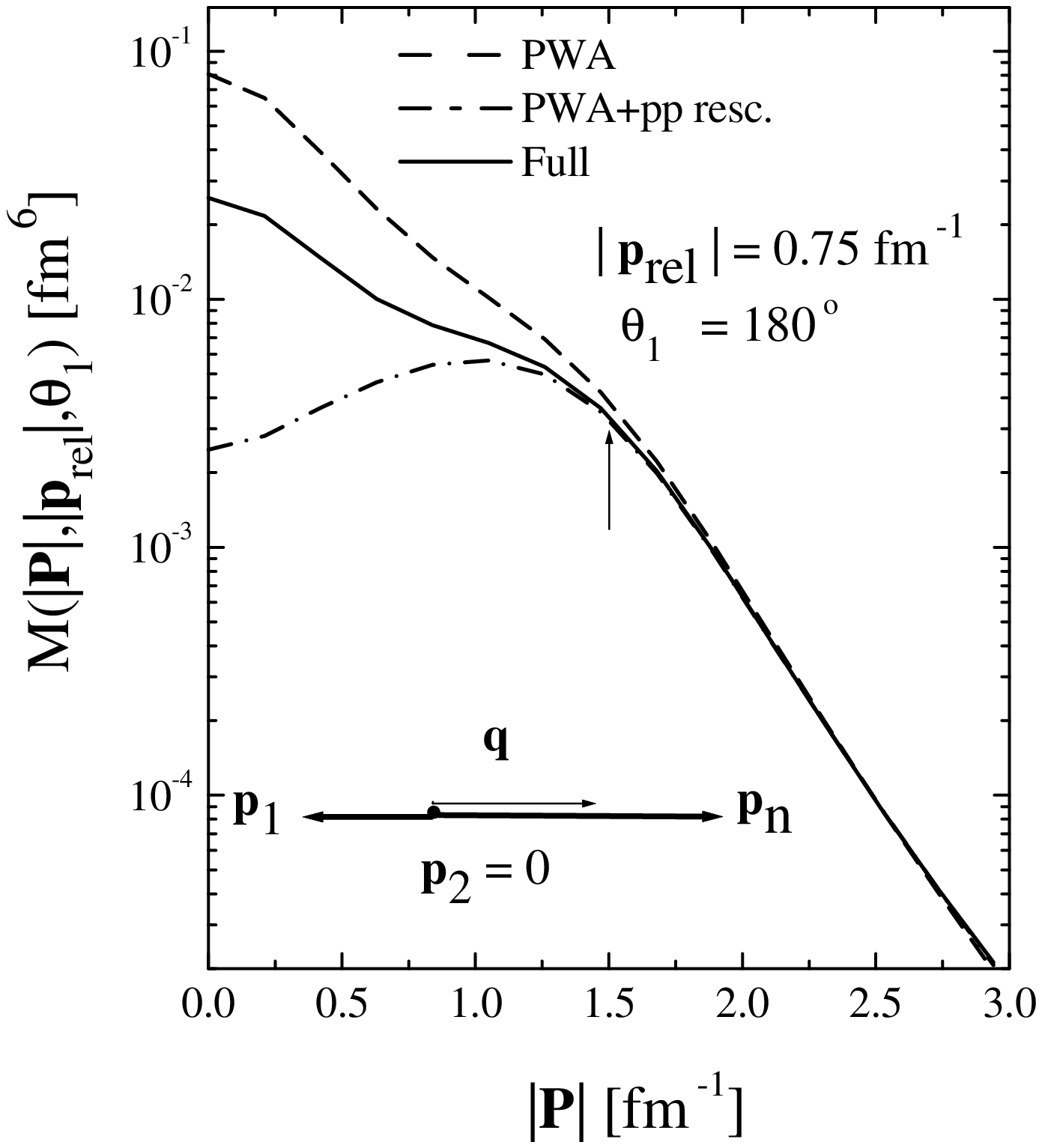}
  \caption{The effects of the FSI on the  $^3He(e,e^\prime p p)n$ process. The  
   transition form factor
      $M(|{\bf p}_n|,|{\bf p}_{rel}|,\theta_1)$ (eq. \ref{emme01}), where  ${\bf p}_n$ is the neutron momentum,
      ${\bf p}_{rel}=({\bf p}_1-{\bf p}_2)/2$ the two-proton relative momentum and
       $\theta_1$ = $\widehat{{\bf p}_n\cdot{\bf p}}_{rel}$, is shown {\it vs}
         the missing momentum   ${\bf P}={\bf p}_1+{\bf p}_2={\bf q}-{\bf p}_n={\bf p}_{m}$\, for fixed values of
      $|{\bf p}_{rel}| = 0.75 fm^{-1}$ \,and \,  $\theta_1 = 180^o$ (super-parallel kinematics). {\bf  PWA}: 
      all particles in the final state are described by plane waves; {\bf PWIA}:  plane wave for the hit neutron
      plus $p-p$ rescattering; {\bf full}:  full three-body rescattering  taken
      into account. The arrow and the momentum vector balance, which refer
      to the dashed and  dot-dashed lines, denote the point corresponding to the
       Two Nucleon Correlation  configuration originating the  bumps in the Spectral Function
       at $ {\bf k}_n^2/(4M_N) \simeq {\bf p}_{rel}^2/M_N$; thus, in the point denoted by the arrow one has
      $|{\bf k}_n|=2|{\bf p}|_{rel}\simeq 1.5$ $fm^{-1}$, 
      ${\bf k}_1\simeq -{\bf k}_n$, and  ${\bf k}_2\simeq 0$. For
      ${\bf P}>1.5$ $fm^{-1}$, the ground state momentum balance is
      always similar to the 2NC configuration (${\bf k}_n\simeq {\bf k}_1$,
      ${\bf k}_2<<{\bf k}_1$), whereas for ${\bf P}<1.5$ $fm^{-1}$,
      the configuration is far from the 2NC one. AV18 interaction \cite{av18}). Three-nucleon wave function
      from \cite{pisa}. (After \cite{claleo}).}\label{fig6}
\end{figure}  
\be
\psi^f({\bf r}_1,{\bf r}_2,{\bf r}_3)\,=\,S^{FSI}({\bf r}_1,{\bf r}_2,{\bf r}_3)
      \,e^{-\,i\,{\bf p}_n\cdot{\bf r}_3}
      \,\phi_{{\bf p},{\bf p}_2}({\bf r}_1,{\bf r}_2)
\ee
where $S^{FSI}$, which is given by eq. (\ref{esse}), describes the interaction
of the fast neutron ``3'' with protons ``1'' and ``2'', whose relative wave function
$\phi_{{\bf p},{\bf p}_2}({\bf r}_1,{\bf r}_2)$ is the solution of
the two-body Schr\"odinger equation in the continuum.
The eight-fold cross section has the following form
\be
\frac{d^8\sigma}{d E_{e^\prime}\,d\Omega_{e^\prime}
      \,d\Omega_{{\bf p}_n}\,d|{\bf p}_{rel}|\,d\Omega_{{\bf p}_{rel}}}\,=\,
      K(Q^2,\nu,{\bf p}_n,{\bf p}_{rel})\, G_E(Q^2)^2\,M({\bf p}_n,{\bf p}_{rel})
\ee
where
\be
\label{emme01}
M({\bf p}_n,{\bf p}_{rel})=M(|{\bf p}_n|,|{\bf p}_{rel}|,\theta_1)
=\left|\int \psi^f({\bf r}_1,{\bf r}_2,{\bf r}_3)
      \,e^{-\,i\,{\bf p}_m\cdot{\bf r}_n}
      \,\psi_{^3He}({\bf r}_1,{\bf r}_2,{\bf r}_3)
      \,\delta(\sum^3_1 {\bf r}_i)\,\prod^3_{i=1} d{\bf r}_i\right|^2
\ee
is the transition form factor, ${\bf p}_{rel}=({\bf p}_1-{\bf p}_2)/2$
the relative momentum of the two protons,
${\bf p}_{m}={\bf q}-{\bf p}_n={\bf p}_1+{\bf p}_2$,
the missing momentum and $\theta_1$ the angle between ${\bf p}_{rel}$ and
${\bf p}_m$ (as before the spin-isospin variables are not explicitly
shown, otherwise a summation over them should appear). The main aim of
Ref. \cite{claleo} was to analyse the problem  as to whether proper kinematical
conditions could be found where the effects of the FSI are minimized.
The results of calculations are shown in Fig. \ref{fig6}, where
the transition form factor calculated in the super-parallel kinematics
($\theta_1=180^o$) is exhibited; in such a kinematics the momenta of 
the three particles in the continuum lie on the same line. In Fig. 
\ref{fig6} the arrow and the momentum vector balance correspond to the
kinematical point when ${\bf k}_1=-{\bf k}_n$, ${\bf k}_2=0$
(or ${\bf p}_1+{\bf p}_n={\bf q}$, ${\bf p}_2=0$, having denoted by
${\bf k}$ $({\bf p})$  nucleon  momenta before (after) $\gamma^\star$
absorption).
The  point, in which $|{\bf k}_n|=2|{\bf p}_{rel}|$, corresponds to
the two-nucleon correlation (2NC) configuration originating the bumps 
in the neutron Spectral Function at $E_m=\frac{{\bf k}^2}{4\,M_N}$,
shown in Fig. \ref{fig5}. It can be seen from Fig. \ref{fig6} that right
to the 2NC point, the effects from the  FSI ($n-(pp)$) and ($(p-p)$) is
irrelevant.\\

\centerline{\textbf{The process  $^4He(e,e^\prime p)^3H$}}
\vskip 2mm

Recently \cite{bra01} the effects of color transparency in
quasi-elastic lepton scattering off nuclei have been introduced
by explicitly considering the finite formation time (FFT) that the
hit hadron needs to evolve to its asymptotic physical state.
Within the eikonal approach  the cross section for 
the process $^4He(e,e^\prime p)^3H$ will depend upon the distorted momentum 
distributions
\begin{eqnarray}
n_D({\bf p}_m)  &=& \left |(2\pi)^{-3/2}\int d{\bf r} \,  
\exp(-i{\bf p}_m\cdot{\bf r})I({\bf r})\right |^2
\end{eqnarray}
where  $I({\bf r})$ denotes the distorted overlap between the 
ground state wave functions of nuclei $A$ and $(A-1)$, {\it viz} ( $\xi$ denotes the proper set of Jacobi coordinates)
\begin{figure}[!htp]                  
\centerline{
      \epsfysize=6.5cm\epsfxsize=8cm\epsfbox{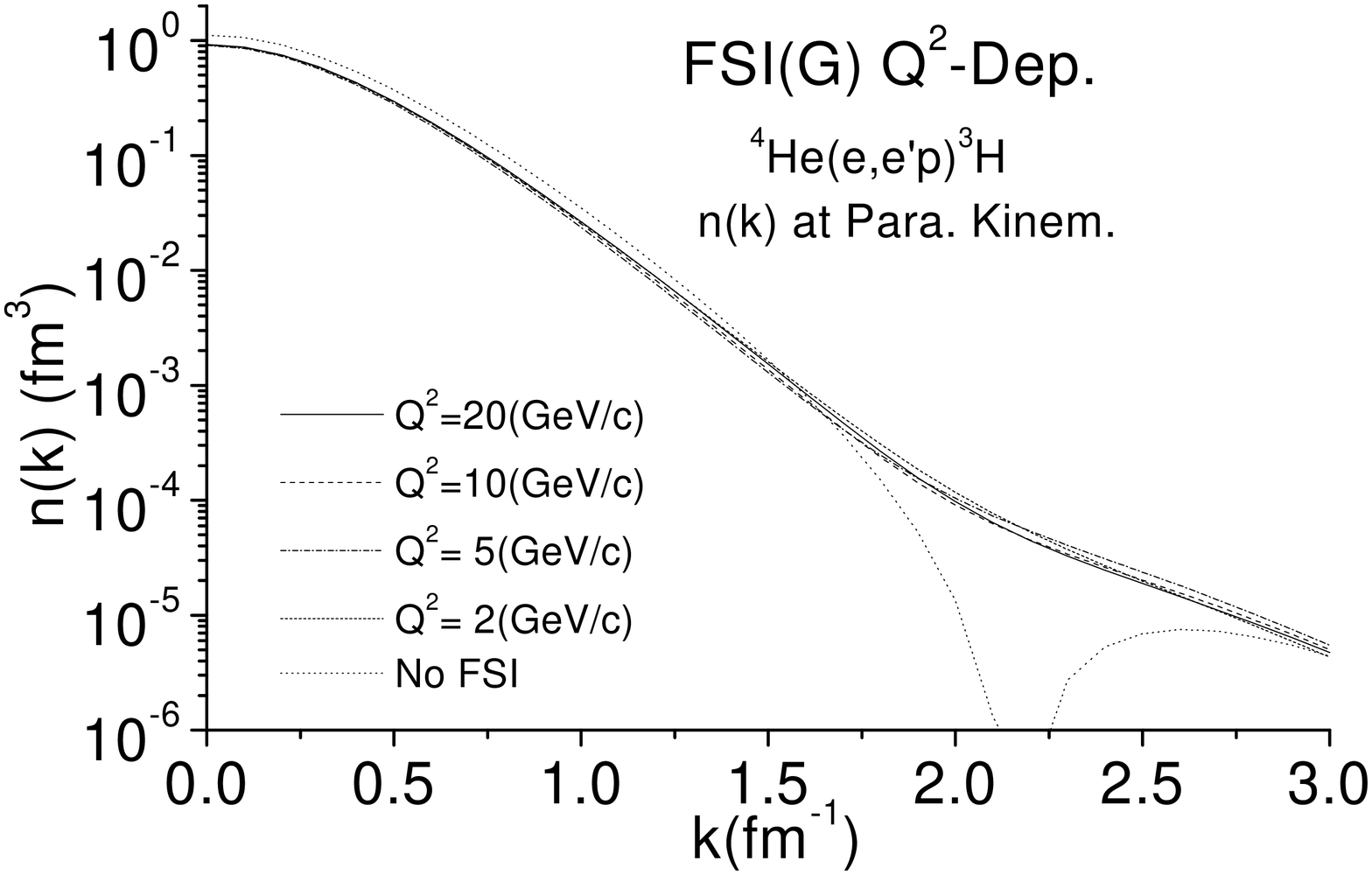}
      \hspace{1mm}
      \epsfysize=6.5cm\epsfxsize=8cm\epsfbox{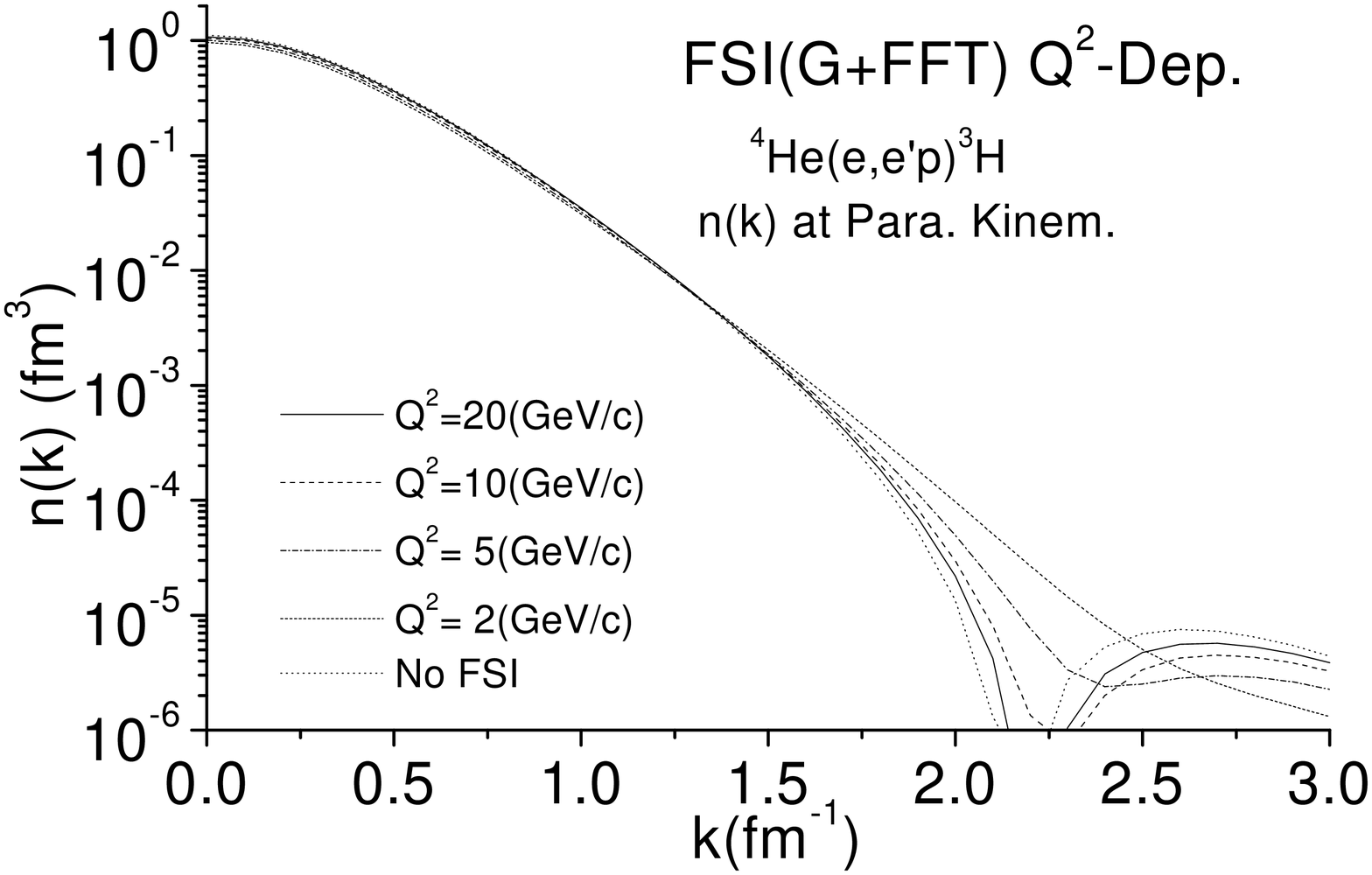}
      }
\caption{{\it Left panel}: the $Q^2$ dependence of FSI effects
calculated within the Glauber approach. {\it Right panel}:  same as in the 
left panel but with the  FFT taken into account. Both Figures
      refer to parallel kinematics $\theta_{\widehat{{\bf q}{\bf p}_m}}=0^o$.
      In this Figure $k \equiv |{\bf p}_m|$. Four-body wave function from \cite{mor02}.
      (After \cite{mor01})}
      \label{fig7} 
\end{figure}
\begin{eqnarray}
 I({\bf r})     &=& \sqrt{A}\int 
   \psi_{A-1}^*(\xi_{A-1}){\cal S}^{FSI}
   \psi_{A}(\xi_{A})\prod_{\xi_i}d\xi_i.
\end{eqnarray}
\noindent  and 
\begin{eqnarray}
{\cal S}^{FSI}({\bf r}_1,{\bf r}_2,{\bf r}_3,{\bf r}_4)   = \prod_{i=1}^{A-1} G(Ai),  \qquad 
G(Ai) = 1-\theta(z_A-z_i)\Gamma({\bf b}_A-{\bf b}_i), 
\label{eq:Glauber}
\end{eqnarray}
is the usual Glauber operator (the hit nucleon is labelled by $A$).
When FFT effects are considered, the $G(Ai)$ can be replaced by \cite{bra01}
\begin{eqnarray}
G(Ai)       = 1-\mathcal{J}(z_i-z_A)\Gamma({\bf b}_A-{\bf b}_i), \qquad 
\mathcal{J}(z) = \theta(z) \left (1-exp\left[-\frac{zxM_NM^2}{Q^2}\right]\right ), 
\label{eq:FFT}
\end{eqnarray}
where $x$ is  the Bjorken scaling  variable, $m$ the  nucleon mass, and  $M$
represents the  average virtuality defined by $M^2 = (m^*_{Av})^2 - M_N^2$.
Eq. \ref{eq:FFT} shows that at high values of  $Q^2$
FFT effects reduce the  Glauber-type FSI, depending on the value of  $M$.
In Ref. \cite{mor01} the value of the average excitation
mass  $m^*_{Av}$ was  taken to be $1.8(GeV/c)$ \cite{bra01}.
the results of calculations are shown in Fig. \ref{fig7}. It can be seen
that the exclusive process $^4He(e,e^\prime p)^3H$ at high values of $Q^2$
could provide a clear cut check of various models which go beyond the
treatment of FSI effects in terms of Glauber-type rescattering.
A clean and regular $Q^2$ behaviour leading to the vanishing of FSI effects
at moderately large values of $Q^2$ is predicted and could be
validated by the experimental observation of a dip in the cross 
section at $p_m \simeq 2.2 fm^{-1}$. Recently, Benhar {\it et al}
\cite{Benhar00} have analyzed the same process, 
viz. the  $^4He(e,e^\prime p)^3H$ reaction, using a colour transparency model.
At variance with the results of Ref. \cite{mor01}, their model does not lead to the vanishing of
FSI at  $Q^2 \simeq 20(GeV/c)^2$. Therefore, it appears that  exclusive 
electron scattering off $^4$He at high $Q^2$ would really represent a powerful tool to  
discriminate various models of hadronic final state rescattering.


\section{Complex nuclei (${\bf Refs.}$ \cite{alv01,alv02,alv03})}
\label{sec:3}

\subsection{Cluster expansion and the nuclear wave function}
\label{subs:3a}

In the  linked-cluster expansion approach developed in Ref. \cite{alv01,alv02,alv03}, the expectation value of a
certain operator $\hat{\mathcal{O}}$ 
\begin{equation}
\label{omedio1}
\langle\hat{\mathcal{O}}\rangle\,=\,\frac{\langle\Psi_A|\,\hat{\mathcal{O}}
\,|\Psi_A\rangle}{\langle\Psi_A | \Psi_A\rangle}
\end{equation} 
is evaluated with correlated wave functions of the following "classical" form
\begin{equation}
\label{psi1}
\Psi_A\,=\,\hat{F}({\bf r}_1,...,{\bf r}_A)\,\Phi_A({\bf r}_1,...,{\bf r}_A)\,,
\end{equation} 
where $\Phi_A$ is a mean field (Slater determinant) wave function, and  $\hat{F}$
a symmetrized (by the symmetrization operator $\hat{S}$) correlation
operator which generates $\textit{correlations}$ into the mean field wave function;
it  has the following general form
\begin{equation}
\label{corre1}
\hat{F}\,=\,{{\hat{S}}}\prod^A_{i<j}\hat{f}(r_{ij}) \,\,
\end{equation}
with
\begin{equation}
\hat{f}(r_{ij})=\sum_p\,f^{(p)}(r_{ij})\,\hat{O}^{(p)}_{ij}
\label{effecorr}
\end{equation}
where the operators $\hat{O}^{(p)}$ are the same which appear in the 
two-nucleon interaction, having the form ( e.g. in case of  a $V8$-type interaction) 
\begin{equation}
\label{operator}
{\hat{O}}^{p=1-8}_{ij}=\left[1,\, {\bf \sigma}_i
      \cdot{{\bf \sigma}}_j,\,S_{ij},\,({\bf L} \cdot {\bf S})_{ij}\right]\otimes
      \left[1,\,{\bf \tau}_i\cdot\ {\bf \tau}_j \right]\,
\end{equation}
The central parts $f^{(p)}(r_{ij})$ of the correlation function
${\hat f}^{(p)}$, reflect the radial behaviour of the various components
and their actual form is determined either by the minimization of the ground
state energy, or by other criteria. 

The cluster expansion of Eq.\ref{omedio1} is carried out in terms of the
quantity $\hat{\eta}_{ij}={\hat{f}}^2_{ij}-1$,  whose integral plays the
role of a small expansion parameter; by expanding the numerator and the
denominator  the terms $\hat{\mathcal{O}}_n$  of the same order $n$ in
$\eta_{ij}$, are collected obtaining $\langle\hat{\mathcal{O}}\rangle=
\mathcal{O}_0+\mathcal{O}_1+\mathcal{O}_2+...$, with
\begin{eqnarray}
\label{eta1}
\mathcal{O}_0&=&\langle\hat{\mathcal{O}}\rangle\, \nonumber\hspace{2.73cm}
      \mathcal{O}_1\,=\,\langle\sum_{ij}
      \hat{\eta}_{ij}\,\hat{\mathcal{O}}\rangle\,
      -\,\mathcal{O}_0\,\langle\sum_{ij}\,
      \hat{\eta}_{ij}\rangle\,\nonumber\\
\nonumber\\
\mathcal{O}_2&=&\langle\sum_{ij<kl}\hat{\eta}_{ij}\,\hat{\eta}_{kl}\,
      \hat{\mathcal{O}}\rangle\,
      -\,\langle\sum_{ij}\hat{\eta}_{ij}\,\hat{\mathcal{O}}\rangle\,
      \langle\sum_{ij}\hat{\eta}_{ij}\rangle\,+   
\mathcal{O}_0\,\left(\langle\sum_{ij<kl}\hat{\eta}_{ij}\,
      \hat{\eta}_{kl}\,\rangle\,-\langle\sum_{ij}\hat{\eta}_{ij}\rangle^2
      \right)\,;
\end{eqnarray}
where $\langle [...] \rangle \equiv \langle \Phi_A \left|[...] 
\right|\Phi_A \rangle$. In Refs. \cite{alv01,alv02,alv03}, the 
ground-state energy was minimized  at various orders of the cluster expansion
and the parameters characterizing the correlation functions and 
the mean-field single-particle wave function have been used in the calculation
of the the transition matrix 
elements of various  
electro-disintegration processes using the same cluster expansion 
employed to calculate the energy. The ground 
state energy of $^{16}O$ and $^{40}Ca$ has been calculated using the Argonne 
$V8^\prime$ potential \cite{pud01}  and adopting, as in Ref. 
\cite{fab01}, the so called $f_6$ approximation consisting in 
considering only the first six components of Eq. \ref{effecorr}. 
The expectation value of the  Hamiltonian 
has been  obtained by calculating the average values of 
the kinetic and potential energies, i.e. 
\begin{equation}
\label{kin1}
\langle\hat{T}\rangle\,=\,-\frac{{\hbar}^2}{2m}\,\int\,d{\bf k}\,k^2\,n(|{\bf k}|)\,,
\end{equation}
where $n({\bf k})$ is the nucleon momentum distribution, 
\begin{equation}
\label{momdis1}
n(|{\bf k}|)\,=\,\frac{1}{(2\pi)^3}\,\int\,d{\bf r}_1\,d{\bf r}_1^\prime
\,e^{-i\,{\bf k}\cdot({\bf r}_1-{\bf r}^\prime_1)}\,\rho^{(1)}
      ({\bf r}_1,{\bf r}^\prime_1),
\end{equation}
and
\begin{equation}
\label{pot1}
\langle\hat{V}\rangle\,=\,\frac{1}{2}\sum_{i<j}\langle \hat {v}_{ij}\rangle
=\,\frac{A(A-1)}{2}\sum_p\;\int\;d{\bf r}_1 d
      {\bf r}_2\;v^{(p)}(r_{12})\rho^{(2)}_{(p)}({\bf r}_1,{\bf r}_2)\,.
\end{equation}
The calculations have been performed by cluster expanding the expectation 
value of the non  diagonal one-body, $\hat{\rho}^{(1)}$, and diagonal
two-body, $\hat{\rho}^{(2)}({\bf r}_1,{\bf r}_2)$, density matrix operators. 
The six correlation functions $f^{(p)}(r_{ij})$ were the ones obtained in 
Ref. \cite{fab01}, and the mean field motion has been described by   Harmonic Oscillator (HO) and Saxon-Woods (SW)
{\it single particle wave functions} (spwfs). 
\begin{figure}[!hbp]                        
\centerline{
      \epsfysize=0.47\textwidth\rotatebox{-90}{\epsfbox{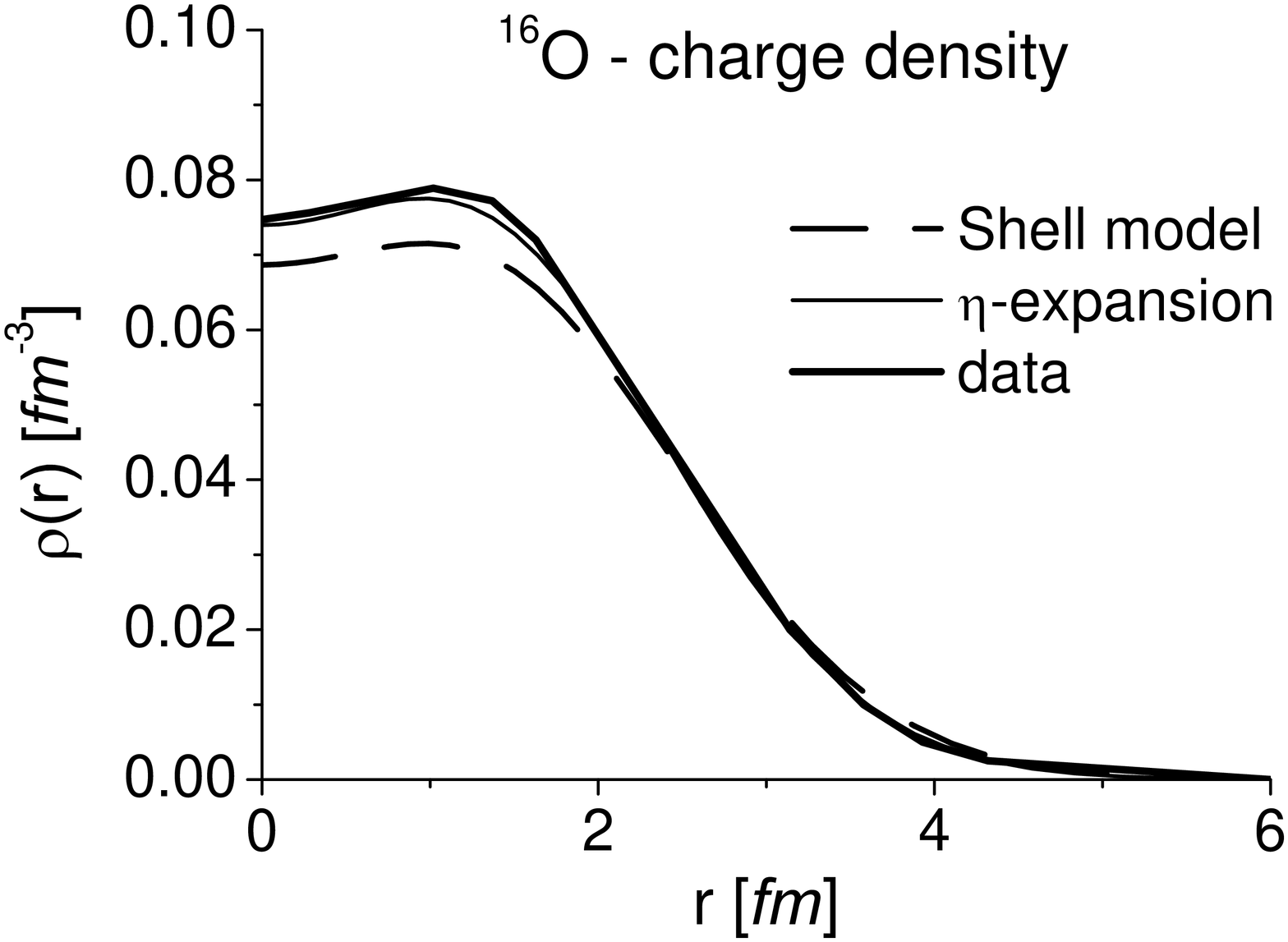}}
      \hspace{1mm}
      \epsfysize=0.47\textwidth\rotatebox{-90}{\epsfbox{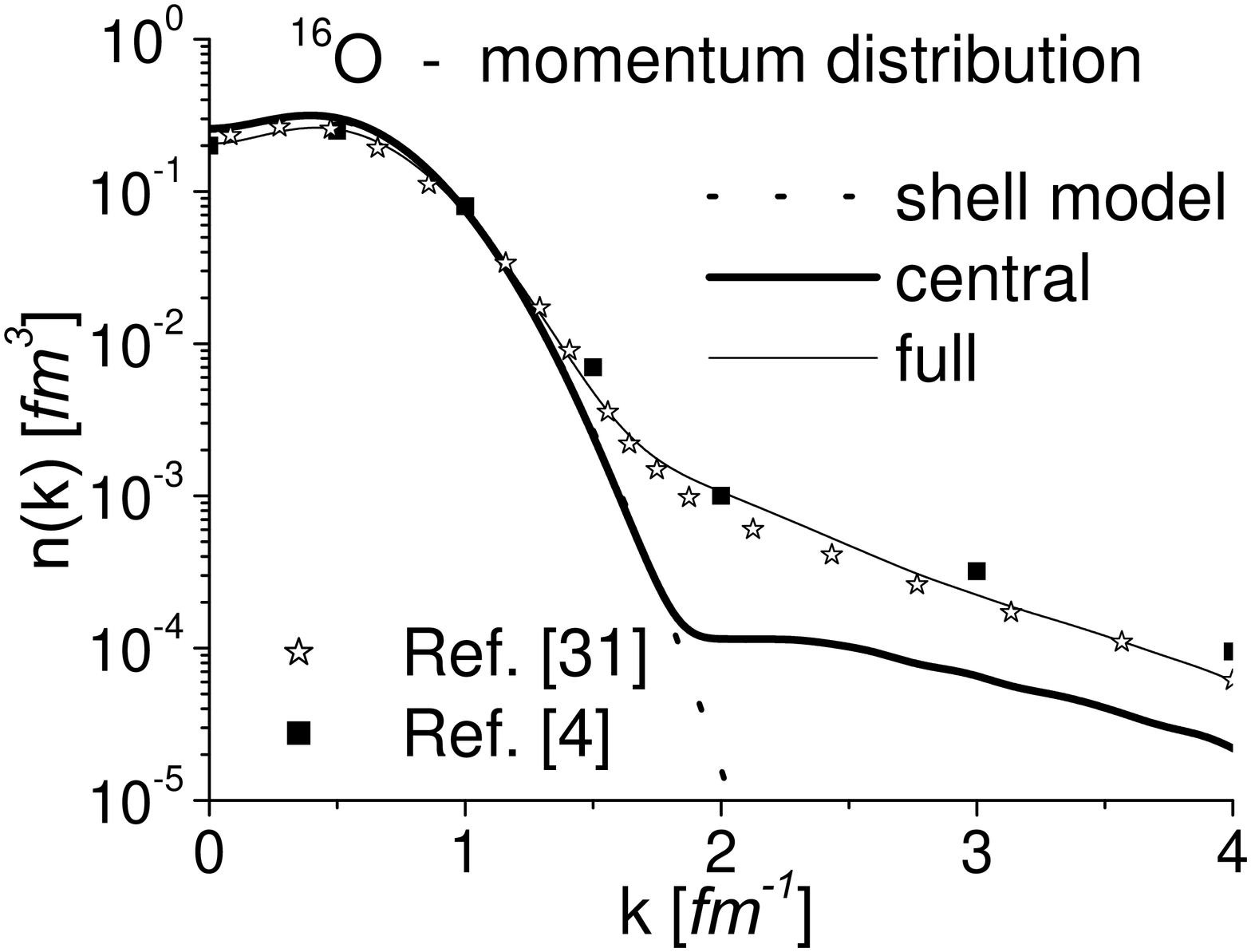}}}
\caption{{\it Left panel}: the  charge density 
      of $^{16}O$ calculated using the cluster expansion (\ref{eta1}) with Harmonic Oscillator 
      (HO) \textit{spwf}. \textit{Dashed line}: mean-field wave results;
       \textit{full line}: results of the cluster expansion; \textit{thick full line}: 
      experimental data \cite{dev01}. {\it Right panel}: the momentum distribution of $^{16}O$. 
      \textit{Dotted line}: mean-field result; 
      \textit{full line}: full correlated result; \textit{thick full line}: central correlation only; 
      \textit{open stars}: the FHNC result \cite{fab01}; \textit{full squares:} the VMC result \cite{pie01}.
      The normalization of the density is $4\pi\int \rho(r)\,r^2\,dr = Z$, $Z$
      being the number of protons, and the normalization of $n(k)$ is 
      $4 \pi \int n(k)\,k^2\,dk\,=\,1$. (After \cite{alv01})}
      \label{fig8}
\end{figure}
\begin{figure}[!hbp]                        
\centerline{
      \epsfysize=0.47\textwidth\rotatebox{-90}{\epsfbox{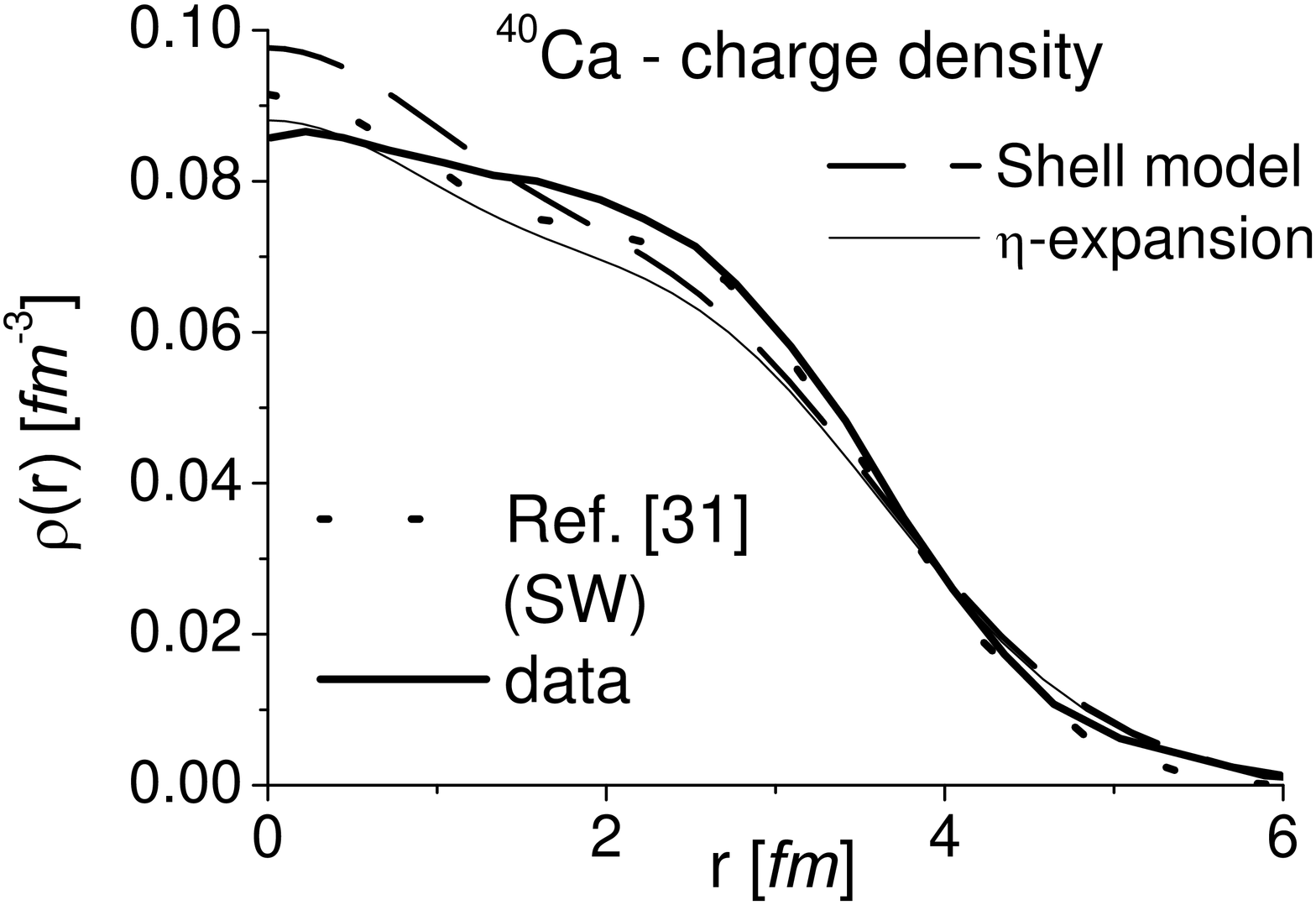}}
      \hspace{1mm}
      \epsfysize=0.47\textwidth\rotatebox{-90}{\epsfbox{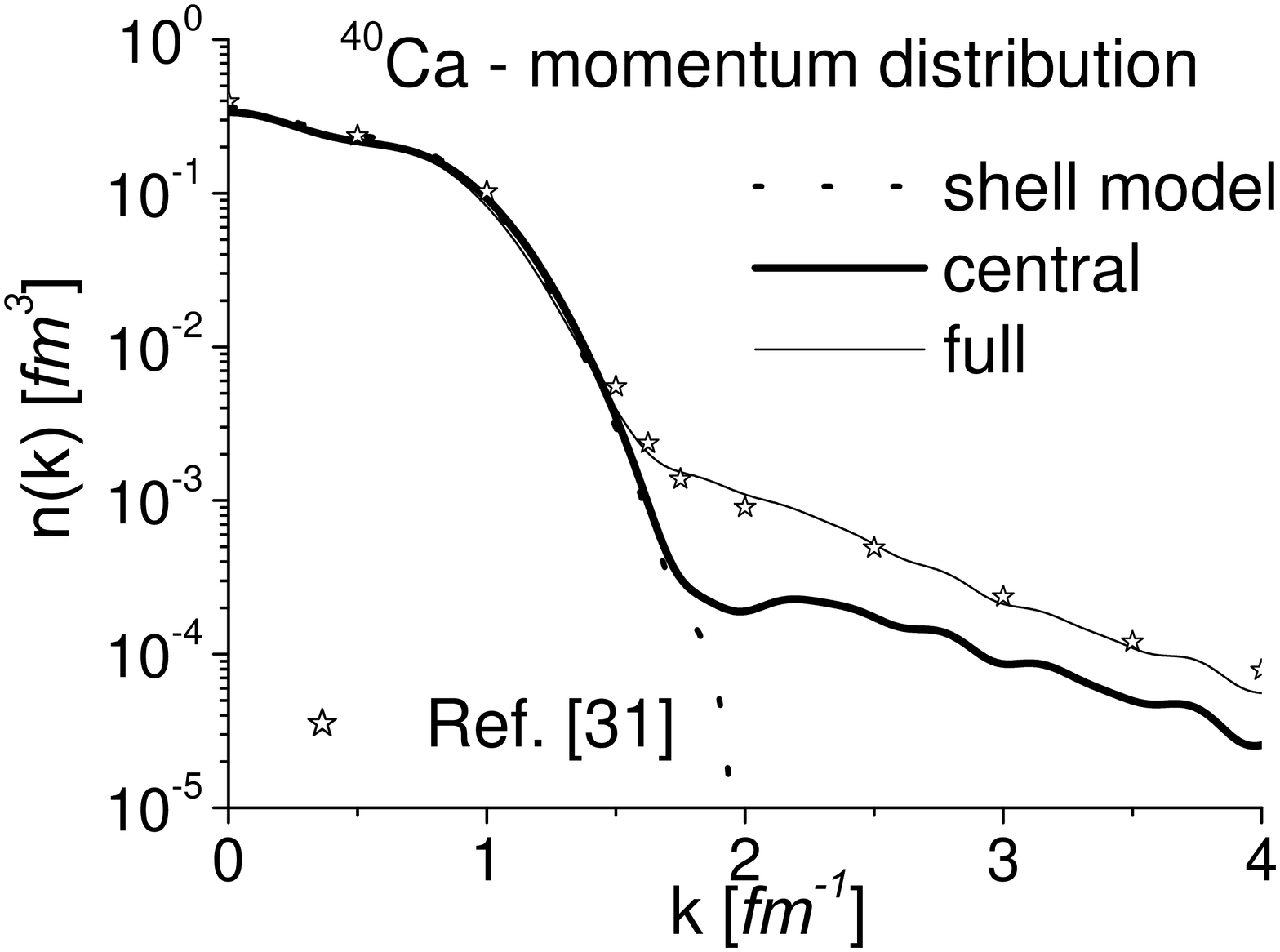}}}
\caption{The same as in Fig. \ref{fig8}, for $^{40}Ca$. In the left panel, the additional dotted line
      was  obtained in Ref. \cite{fab01} with SW spwfs.}
\label{fig9}
\vskip -0.5cm
\end{figure}
\begin{table}[!htp]
   \caption{The  potential, kinetic and total energies per particle   for 
   $^{16}O$  calculated at 1-st order of the  $\eta$-expansion ($\mathcal O_0 + \mathcal O_1$ in Eq. \ref{eta1}).
    The correlation function and the values of the  harmonic oscillator (HO)
      and Saxon-Woods (SW) parameters are the same as in Ref. \cite{fab01}.
       The results from the latter are listed in brackets. (After Ref. \cite{alv03})}
    \begin{center}
      \begin{tabular*}{\textwidth}{p{4cm}p{3cm}p{3cm}p{3cm}}
   & $\langle T\rangle/A$ (MeV) & $\langle V\rangle/A$ (MeV) & $E/A$ (MeV)\\\hline
HO & 22.4 (22.6) & -26.6 (-27.4) & -4.2 (-4.9)\\
SW & 28.4 (27.3) & -31.5 (-32.4) & -3.1 (-5.1)\\\hline
      \end{tabular*}
   \end{center}
   \label{ossigeno}
\vskip -0.5cm
\end{table}
The ground-state  energy, the density and the momentum distribution for 
$^{16}O$ and $^{40}Ca$ have been calculated at  first order of the $\eta$-expansion and, 
as in Ref. \cite{fab01}, it has been found that the charge densities
corresponding to the minimum of the energy, appreciably disagree with the
corresponding experimental quantities; thus, in line with Ref. \cite{fab01},  the mean-field parameters 
have been changed to obtain agreement between
theoretical and experimental charge densities; such a procedure is justified by the mild  dependence
of  the energy around the minimum.
Typical results for the energy are shown in Table \ref{ossigeno}. It should be pointed out that the 
value of the contribution from the seventh and eighth components of the potential have only be estimated. 
The results for the charge densities and momentum distributions
are shown in Figs. \ref{fig8} and \ref{fig9}. It can be seen that 
the agreement between the  results of the cluster expansion from \cite{alv01,alv02,alv03} and the FHNC/SOC from
\cite{fab01} appears to be a  very satisfactory one.  Both  approaches predict momentum distributions which 
do not appreciably differ from the ones obtained in Ref. \cite{pie01}, where
the Variational Monte Carlo method and the $AV18$ interaction have been used;
the dominant non-central correlations are the isospin,
$f_4=f^{(4)}(r_{ij}) {\bf \tau}_i\cdot{\bf \tau}_j$, and isospin-tensor,  
$f_6=f^{(6)}(r_{ij}) {\bf \tau}_i\cdot{\bf \tau}_j S_{ij}$, correlations.
In order to investigate the convergence properties of the momentum distribution
the  second order cluster contribution to the momentum distribution of  $^{16}O$
has been evaluated (\cite{alv03});
the results are  shown in Fig. \ref{fig10} and it can be seen that the
convergence is very good. 
The convergence of the energy is being investigated, also by introducing a new cluster expansion 
which effectively includes higher order terms (see Ref.\cite{alv01}).
A preliminary comparison with the results from Ref. \cite{fab02}
 indicates that the value of the kinetic energy for $^{40}Ca$ is stable,  which makes us confident
that also  the momentum distributions of $^{40}Ca$  converge at 1st order.
\begin{figure}[!hbp]                       
\centerline{
      \epsfysize=9cm\epsfxsize=6.5cm\rotatebox{-90}{\epsfbox{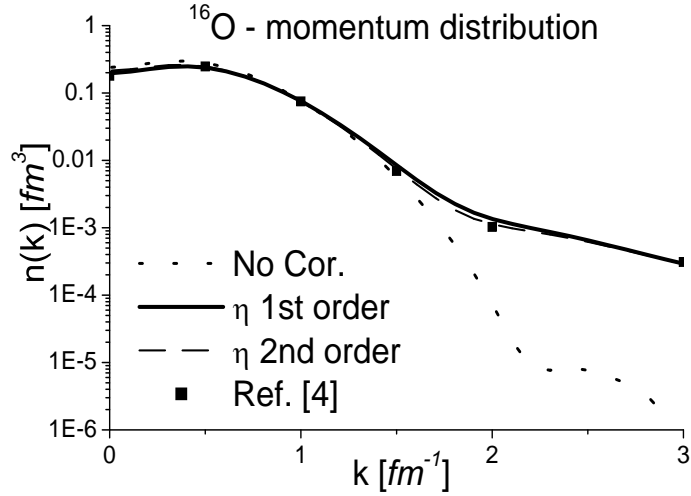}}}
\hspace{12cm}\caption{The nucleon momentum distribution in $^{16}O$ momentum distributions calculated up to second
      order in the $\eta$-expansion using Saxon-Woods spwfs and
      the $f_1$, $f_4$ and $f_6$ correlation functions from \cite{pie01}.
      The dotted line represents the mean field result, whereas the dashed (full) line is the result of the 
      cluster expansion at 1st (2nd) order given by  $\mathcal O_0 + \mathcal O_1$ 
      ($\mathcal O_0 + \mathcal O_1 + \mathcal O_2$) in Eq. \ref{eta1}. 
      (After \cite{alv01})}
\label{fig10}
\vskip -0.5cm
\end{figure}
\subsection{FSI in $A(e,e^\prime p)X$: the Glauber approach}
\label{subs:3b}
 
The semi-inclusive $A(e,e^\prime p)X$ process denotes the process in which a summation over all excited states of $(A-1)$, or,
equivalently, over the missing energy $E_m = M_N + M_{A-1} - M_A + E_{A-1}^*$, 
has been carried out. The cross section 
(\ref{eq2})   becomes then  proportional to the \textit{distorted} 
distorted momentum distributions (see e.g. \cite{niko})
\begin{equation}
n_D({\bf p}_m)={(2 \pi)^{-3}} \int e^{i {\bf p}_m({\bf r}_1 -{\bf r}^\prime_1)}
      \rho_D ({\bf r}_1,{\bf r}^\prime_1) d{\bf r}_1 d{\bf r}^\prime_1
\label{nd}
\end{equation}
where 
\begin{eqnarray}
\rho_D ({\bf r}_1,{\bf r}^\prime_1)= \frac {\langle\Psi_A\,|\,{S^{FSI}}^{\dagger}
\,\hat{O}({\bf r}_1,{\bf r}^\prime_1)\,{S^{FSI}}^\prime\,|\,{\Psi_A}^\prime\rangle}
      {\langle\Psi_A\,|\,\Psi_A\rangle}
\label{rodi}
\end{eqnarray}
\begin{figure}[!hpt]                                              
\centerline{
      \epsfysize=0.49\textwidth\rotatebox{-90}{\epsfbox{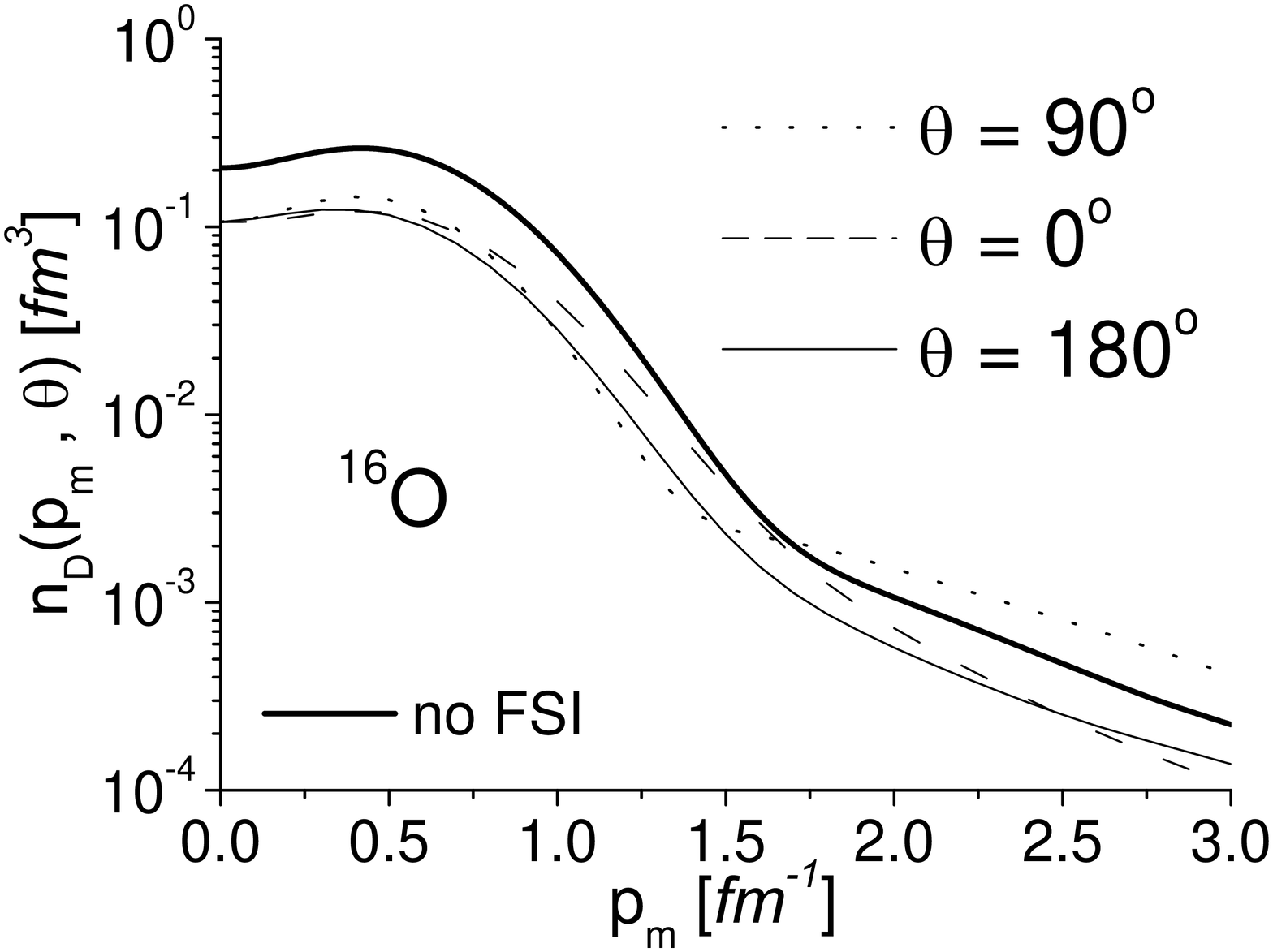}}
      \hspace{1mm}
      \epsfysize=0.49\textwidth\rotatebox{-90}{\epsfbox{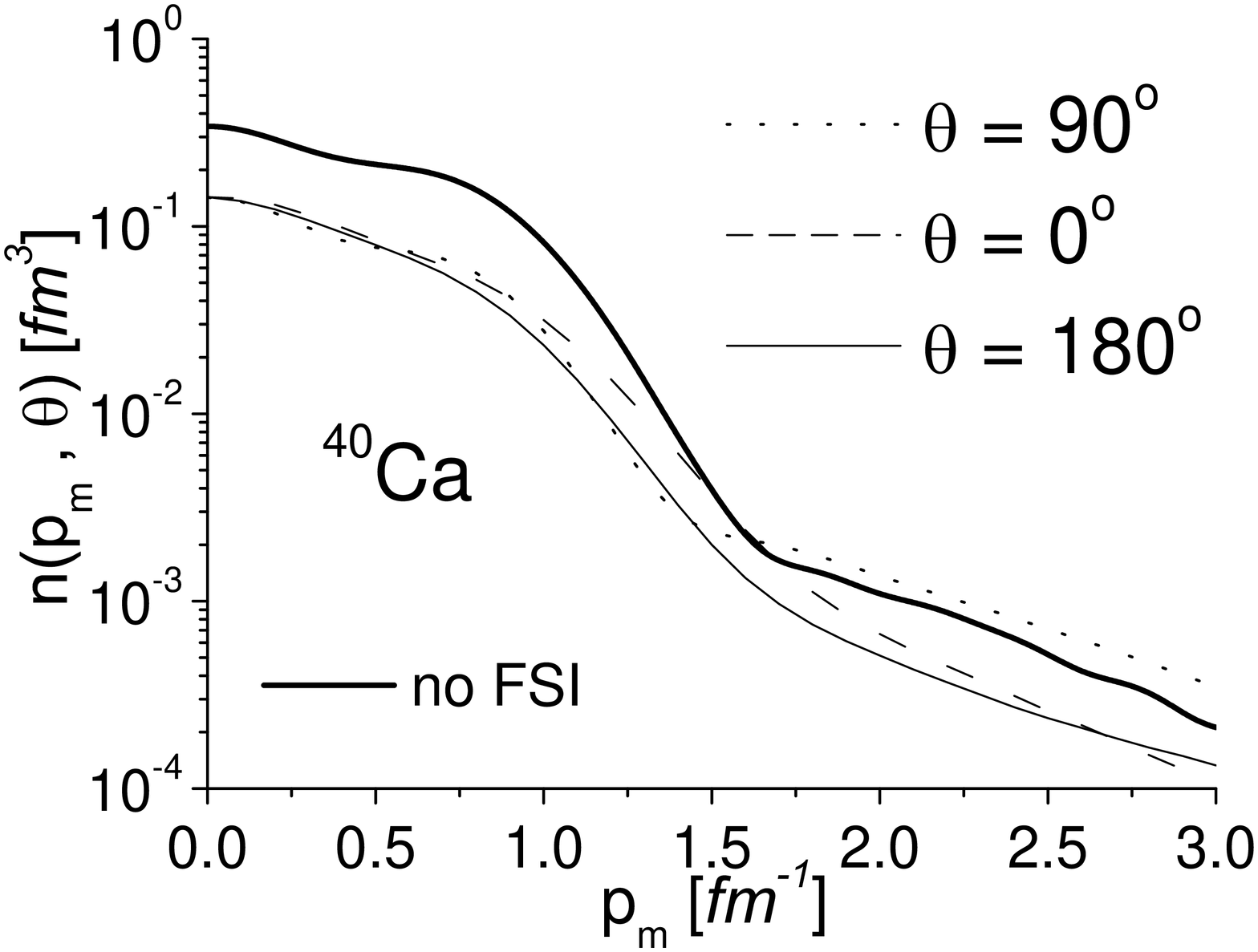}}
      }
\caption{The distorted momentum distribution, $n_D({\bf p}_m)=n_D(p_m, \theta)$ 
      ($\theta=\widehat{{\bf q}{\bf p}}_m$),  for  $^{16}O$  and $^{40}Ca$,
      obtained  from Eq. \ref{nd} using  correlated wave functions, Harmonic Oscillator
      $\textit{spwf}s$ and the Glauber operator  (\ref{SG}). The value of the
      integrated nuclear transparency (\ref{intnd}) for $^{16}O$ is $0.5$.
      (After \cite{alv01})}
      \label{fig11}
\end{figure}
\noindent is the  distorted one-body mixed density matrix, 
$S^{FSI}$  describes the  FSI,  and  the primed quantities have
to  be evaluated at ${\bf r}^\prime_1$, and ${\bf r}_i$, with $i=2, ...,A$.
The {\it nuclear transparency}  $T$ is defined as follows 
\begin{equation}
T = \frac{\int n_D({\bf p}_m) d{\bf p}_m}{\int n(|{\bf k}|) d{\bf k}} = 
      \int \rho_D ({\bf r})d{\bf r} = 1+ \Delta T
\label{intnd}
\end{equation}
where $\rho_D({\bf r})=\rho_D ({\bf r}_1={\bf r}^\prime_1\equiv {\bf r})$
and  $\Delta T$ originates from the FSI.
In Ref. \cite{cio01} Eq. \ref{nd} has been evaluated using a Glauber
representation for the operator  $S$, \textit{viz}
\begin{equation}
S^{FSI}({\bf r}_1,{\bf r}_2, \dots,{\bf r}_A) = \prod_{j=2}^AG({\bf r}_1,{\bf r}_j)
      \equiv \prod_{j=2}^A\bigl[1-\theta(z_j-z_1)\Gamma({\bf b}_1-{\bf b}_j)\bigr]
\label{SG}
\end{equation}
where we remind that ${\bf b}_j$ and $z_j$ are the transverse and the longitudinal components 
of the nucleon coordinate ${\bf r}_j\equiv({\bf  b}_j,z_j)$, ${\mit\Gamma}({\bf b})$
the Glauber profile function for elastic proton nucleon scattering, and the 
function $\theta(z_j-z_1)$ takes care of the fact that the struck proton
``1''  propagates along a straight-path trajectory so that it interacts with
nucleon  ``$j$'' only if $z_j>z_1$. The same cluster expansion described in
Section \ref{subs:2a} has been used to evaluate Eq. \ref{intnd} taking Glauber 
rescattering exactly into account at
the given order $n$, and using the approximation $|\Psi_{A-3}|^2 =
\prod_3^A \rho(i)$.
Using such an approach and the mean-field and correlation parameters obtained from the
energy calculation,  the  \textit{distorted} nucleon momentum
distributions $n_D({\bf p}_m)=n_D(p_m, \theta)$, where $\theta$ is the angle
 between ${\bf q}$ and ${\bf p}_m$, has been obtained in Ref. \cite{alv01,alv02,alv03}; the results for 
   $^{16}O$ and $^{40}Ca$
are presented in Fig. \ref{fig11}.
\subsection{Finite formation time effects}
\label{subs:3c}
As already discussed in Section \ref{subs:2b} the effects of color transparency
in quasi-elastic lepton scattering off nuclei, can be introduced
by considering the finite formation time (FFT) that the
hit hadron needs to evolve to its asymptotic physical state.
Following Section \ref{subs:2b} the Glauber factor $G({\bf r}_1,{\bf r}_j)$ in Eq. \ref{SG}
is replaced by 
\begin{equation}
G({\bf r}_1,{\bf r}_j)= 1-\mathcal{J}(z_j-z_1)\Gamma({\bf b}_1-{\bf b}_j)
\end{equation}
with 
\begin{equation}
\mathcal{J}(z) = \theta(z) \left (1-exp\left[-\frac{zxM_NM^2}{Q^2}\right]\right ), 
\end{equation}
where  $x=Q^2/(2M_N\nu)$  being  the Bjorken scaling 
 variable,
 and 
 $M^2 = (m^*_{Av})^2 - M_N^2$ with   $m^*_{Av} = 1.8(GeV/c)$.
    A suitable quantity for estimating the effect of FSI on the cross section
   is the 
  \textit{forward}-\textit{backward}  
asymmetry, constructed out of
   forward and backward cross sections, 
    \begin{figure}  [!hpb]                      
\centerline{
      \epsfysize=0.45\textwidth\rotatebox{-90}{\epsfbox{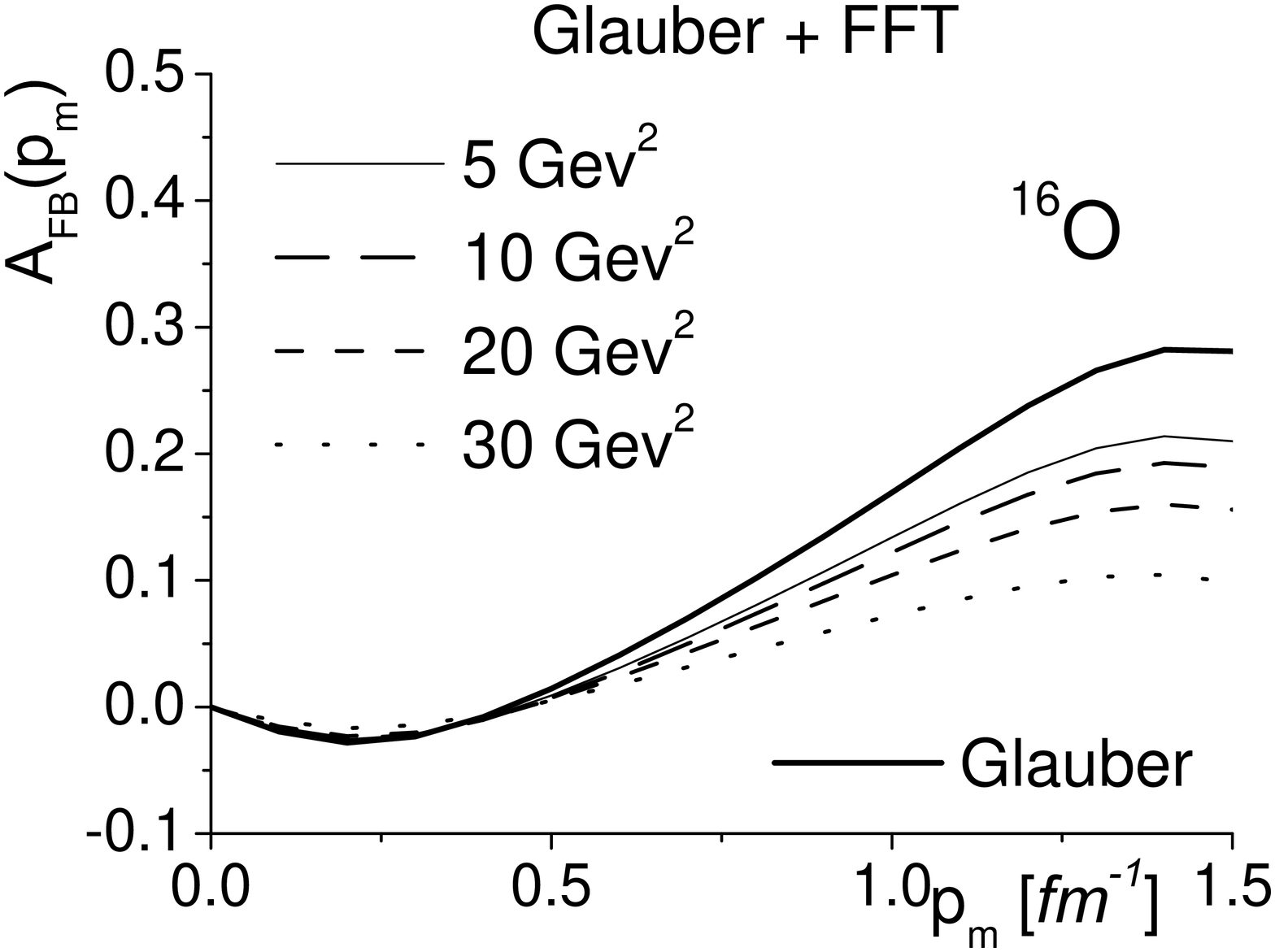}}
      \hspace{1mm}
      \epsfysize=0.45\textwidth\rotatebox{-90}{\epsfbox{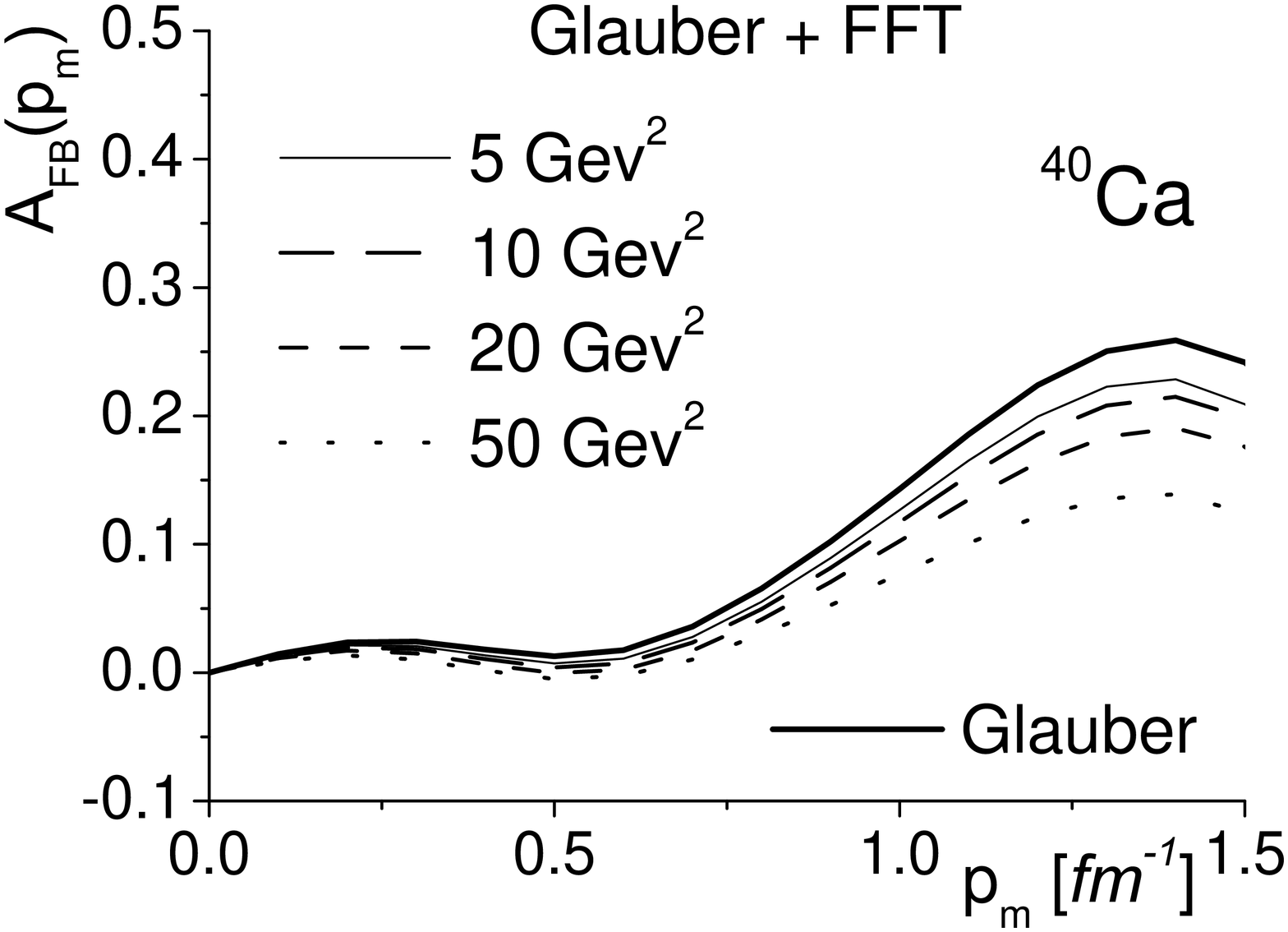}}
      }
\caption{The forward-backward asymmetry defined by eq. (\ref{fba2}),
      for $^{16}O$ (\textit{left}) and $^{40}Ca$ (\textit{right});
      the thick line represents the $Q^2$-independent Glauber result and the other curves include FFT effects.
      (After \cite{alv01})}
\label{fig12}
\end{figure}

\noindent namely
\be
\label{fba1}
A_{FB}(p_m,\theta)\,=\,\frac{\sigma(p_m,\theta=0^o)-\sigma(p_m,\theta=180^o)} 
      {\sigma(p_m,\theta=0^o)+\sigma(p_m,\theta=180^o)}\,
\ee
Within the  factorized approximation  for the cross section, $A_{FB}$ reduces to
 \be
\label{fba2}
A_{FB}(p_m,\theta)\,=\,\frac{n_D(p_m,\theta=0^o)-n_D(p_m,\theta=180^o)} 
      {n_D(p_m,\theta=0^o)+n_D(p_m,\theta=180^o)}\,;
\ee
which obviously vanishes in absence of any FSI. 
The effects of the  FSI and the FSI+FFT on $A_{FB}$ are shown in Fig. \ref{fig12}.
for $^{16}O$ and $^{40}Ca$ for  different values of $Q^2$; it can be seen that
the inclusion of FFT effects strongly affects the forward-backward asymmetry.
\section{Summary and Conclusions}
\label{sec:4}
A realistic parameter-free approach aimed at a consistent
treatment of initial state correlations and  FSI effects in exclusive 
one- and two-hadron emission processes has been developed, which can be applied both to few-
and many-body systems ; in the former case the approach is
based upon the use of realistic few-body wave functions, corresponding to
realistic interactions,  and a generalized eikonal  approach, where the Glauber 
frozen approximation is released. In the case of complex nuclei,  reasonable realistic
 many-body wave functions have been generated by a cluster expansion procedure, which
appears to produce densities and momentum distributions of quality
comparable to  the ones obtained by more advanced many-body approaches.
The  results obtained for both few- and many-nucleon systems seem to  show 
that by a proper choice of the kinematics, FSI effects
might appreciably be reduced, both in one- and two-hadron emissions.
Thus it appears that  by  quasi elastic exclusive processes, the details 
of the ground-state wave function can eventually be investigated.
Preliminary results concerning the generalization of the approach
to take into account color transparency effects have been obtained and work
is in progress in this direction.

\section{Acknowledgments}
The authors are  indebted to  A. Kievsky  for making available  the variational 
three-body  wave functions of the Pisa Group and to Giampaolo C\'o  and Adelchi Fabrocini for providing detailed 
information on the FHNC/SOC   calculation and the numerical values of the correlation functions. 
Useful discussions with M. Braun and  D. Treleani are gratefully acknowledged.
L.P.K. and H. M. are   indebted to  the University of
Perugia and INFN, Sezione di Perugia, for warm hospitality and financial support.
This work was partially supported by the Ministero dell'Istruzione, Universit\`{a} e Ricerca (MIUR), 
through the funds COFIN01.

%
%

\end{document}